%% file: note2469_la.tex
\documentclass[twocolumn,floatfix,showpacs,aps,prd]{revtex4}

\newcommand{\babaryear}       {12}
\newcommand{\babarnumber}     {010}
\newcommand{\slacpubnumber} {15099}

\usepackage[dvips]{graphicx}
\usepackage{epsf} 
\usepackage{epsfig}
\usepackage{rotating}
\usepackage{color}
\usepackage{colordvi}
\usepackage{ifthen}
\usepackage{longtable}
\usepackage{multirow}

\newcommand{\gae}{\lower 2pt \hbox{$\, \buildrel {\scriptstyle >}\over {\scriptstyle \sim}\,$}}
\newcommand{\lae}{\lower 2pt \hbox{$\, \buildrel {\scriptstyle <}\over {\scriptstyle \sim}\,$}}

\RequirePackage{xspace}
\usepackage{relsize}
\def\babar{\mbox{\slshape B\kern-0.1em{\smaller A}\kern-0.1em
    B\kern-0.1em{\smaller A\kern-0.2em R}}}

\def\epem       {\ensuremath{e^+e^-}\xspace}

\def\piz   {\ensuremath{\pi^0}\xspace}
\def\pip   {\ensuremath{\pi^+}\xspace}
\def\pim   {\ensuremath{\pi^-}\xspace}
\def\pimpip  {\ensuremath{\pi^+\pi^-}\xspace}

\def\pipm  {\ensuremath{\pi^\pm}\xspace}

\def\Kp    {\ensuremath{K^+}\xspace}
\def\Km    {\ensuremath{K^-}\xspace}

\def\Kmp   {\ensuremath{K^\mp}\xspace}
\def\KpKm  {\ensuremath{\Kp \kern -0.16em \Km}\xspace}
\def\KS    {\ensuremath{K^0_{\scriptscriptstyle S}}\xspace}

\def\Dbar    {\kern 0.2em\overline{\kern -0.2em D}{}\xspace}

\def\Dz      {\ensuremath{D^0}\xspace}
\def\Dzb     {\ensuremath{\Dbar^0}\xspace}
\def\DzDzb   {\ensuremath{\Dz {\kern -0.16em \Dzb}}\xspace}
\def\Dp      {\ensuremath{D^+}\xspace}

\def\Dstarp  {\ensuremath{D^{*+}}\xspace}

\mathchardef\Upsilon="7107
\def\Y#1S{\ensuremath{\Upsilon{(#1S)}}\xspace}
\def\FourS {\Y4S}

\newcommand{\gev}{\ensuremath{\mathrm{\,Ge\kern -0.1em V}}\xspace}
\newcommand{\mev}{\ensuremath{\mathrm{\,Me\kern -0.1em V}}\xspace}
\newcommand{\kev}{\ensuremath{\mathrm{\,ke\kern -0.1em V}}\xspace}
\newcommand{\gevc}{\ensuremath{{\mathrm{\,Ge\kern -0.1em V\!/}c}}\xspace}
\newcommand{\mevc}{\ensuremath{{\mathrm{\,Me\kern -0.1em V\!/}c}}\xspace}
\newcommand{\gevcc}{\ensuremath{{\mathrm{\,Ge\kern -0.1em V\!/}c^2}}\xspace}
\newcommand{\mevcc}{\ensuremath{{\mathrm{\,Me\kern -0.1em V\!/}c^2}}\xspace}

\def\invfb   {\ensuremath{\mbox{\,fb}^{-1}}\xspace}

\def\ps   {\ensuremath{\rm \,ps}\xspace}

\newcommand{\stat}{\ensuremath{\mathrm{(stat)}}\xspace}
\newcommand{\syst}{\ensuremath{\mathrm{(syst)}}\xspace}

\def\pep2{PEP-II}

\newcommand{\dedx}{\ensuremath{\mathrm{d}\hspace{-0.1em}E/\mathrm{d}x}\xspace}
\newcommand{\chisq}{\ensuremath{\chi^2}\xspace}

\def\CP                {\ensuremath{C\!P}\xspace}
\def\CPV        {\ensuremath{C\!P\!V}\xspace}
\def\CPT               {\ensuremath{C\!PT}\xspace}

\def\geantiv       {\mbox{\tt GEANT 4}\xspace}

\def\DzDzb         {\ensuremath{\Dz\ensuremath{-}\Dzb}\xspace}

\def\pisoftp    {\ensuremath{\pi_{\rm s}^{+}}\xspace}

\def\Kppim      {\ensuremath{K^{+}\pi^{-}}\xspace}
\def\Kmpip      {\ensuremath{K^{-}\pi^{+}}\xspace}

\def\KmKp       {\ensuremath{K^{-}K^{+}}\xspace}
\def\KK         {\ensuremath{K^{-}K^{+}}\xspace}
\def\KpKm       {\ensuremath{K^{+}K^{-}}\xspace}
\def\kk         {\ensuremath{KK}\xspace}
\def\pimpip     {\ensuremath{\pi^{-}\pi^{+}}\xspace}
\def\pipi       {\ensuremath{\pi^{-}\pi^{+}}\xspace}

\def\DzRW       {\ensuremath{\Kmp \pipm}\xspace}

\def\dm         {\ensuremath{\Delta m}\xspace}
\def\t          {\ensuremath{t}}
\def\terr       {\ensuremath{\sigma_{\t}}\xspace}

\def\Pchisq      {\ensuremath{P(\chi^2)}}

\def\Dst         {\ensuremath{D^{*}}}
\def\Dstp        {\ensuremath{D^{*+}}}
\def\Dstm        {\ensuremath{D^{*-}}}

\def\ttrue  {\ensuremath{t_{\rm true}}\xspace}
\def\toff   {\ensuremath{t_0}\xspace}

\def\RterrSig{\ensuremath{H_{\terr}^{\rm sig}}\xspace}

\catcode`\@=11\relax
\newskip\dkwidth
\def\dk{
   \dkwidth=2em plus 0.5 em minus 0.25 em\relax
   {\m@th\mathord{
   \hbox{
      \kern 0.3em
      \raise 0.6ex
      \hbox{
         \vrule width 0.25pt height 0.5\dkwidth depth0pt}
      \kern-1.2pt
      \hbox to 1.1\dkwidth{
         \rightarrowfill}
      \kern0.4em}}
   }
}

\def\rightarrowfill{$\m@th\mathord-\mkern-10mu%
  \cleaders\hbox{$\mkern-2mu\mathord-\mkern-2mu$}\hfill
  \mkern-6mu\mathord\rightarrow$}

\catcode`\@=12\relax

\def\yCP        {\ensuremath{y_{C\!P}}\xspace}

\def\deltaY     {\ensuremath{\Delta Y}\xspace}
\def\Kmpip      {\ensuremath{\Km\pip}\xspace}
\def\pimpip     {\ensuremath{\pim\pip}\xspace}

\def\tauKpi     {\ensuremath{\tau_{K\pi}}\xspace}

\def\tauhhp     {\ensuremath{\tau^+}\xspace}
\def\tauhhm     {\ensuremath{\bar{\tau}^+}\xspace}

\def\Gammahhp     {\ensuremath{\Gamma_{hh}^+}\xspace}
\def\Gammahhm     {\ensuremath{\overline{\Gamma}_{hh}^+}\xspace}

\def\SX         {\ensuremath{S_{F}}\xspace}
\def\Sprime         {\ensuremath{S'_{T}}\xspace}
\def\SKpi       {\ensuremath{S_{K\pi}}\xspace}
\def\SKK        {\ensuremath{S_{KK}}\xspace}
\def\Spipi      {\ensuremath{S_{\pi\pi}}\xspace}
\def\Stag      {\ensuremath{S'_{\rm tag}}\xspace}
\def\Sunt      {\ensuremath{S'_{\rm unt}}\xspace}

\usepackage{relsize}
\usepackage{xspace}

\long\def\inst#1{\par\nobreak\kern 4pt\nobreak
    {\it #1}\par\vskip 10pt plus 3pt minus 3pt}

\begin{document}

\begin{flushleft}
SLAC-PUB-\slacpubnumber\\
\babar-PUB-\babaryear/\babarnumber
\end{flushleft}

\title{
        {\mathversion{bold}
Measurement of $\DzDzb$ Mixing and $\CP$ Violation in
Two-Body $\Dz$ Decays
}}

\input authors_apr2012

\begin{abstract}

  We present a measurement of $\DzDzb$ mixing and $\CP$ violation
  using the ratio of lifetimes simultaneously extracted from
  a sample of $\Dz$ mesons produced through the flavor-tagged
  process $\Dstp \to \Dz \pip$, where $\Dz$ decays to $\DzRW$,
  $\KmKp$, or $\pimpip$, along with the untagged decays
  $\Dz \to \DzRW$ and $\Dz \to \KmKp$. The lifetimes of the
  $\CP$-even, Cabibbo-suppressed modes $\KmKp$ and $\pimpip$ are
  compared to that of the $\CP$-mixed mode $\DzRW$ in order to measure $\yCP$ and $\deltaY$.
  We obtain
  $\yCP     =  [0.72 \pm 0.18 \stat \pm 0.12 \syst]\%$ and
  $\deltaY  =  [0.09 \pm 0.26 \stat \pm 0.06 \syst]\%$,
  where $\deltaY$ constrains possible $\CP$ violation.
  The $\yCP$ result excludes the null mixing hypothesis at $3.3 \sigma$ significance.
  This analysis is based on an integrated luminosity of $468 \invfb$
  collected with the \babar\, detector at the PEP-II asymmetric-energy
  \epem collider.

\end{abstract}


\vfill

\maketitle

\pagestyle{plain}

\section{Introduction}
Several measurements~\cite{Aubert:2007wf,Aubert:2007en,Aubert:2009ai,Staric:2007dt,Abe:2007rd,CDF:2007uc}
show evidence for mixing in the $\DzDzb$
system consistent with predictions of possible Standard Model (SM)
contributions~\cite{Wolfenstein:1985ft,Donoghue:1985hh,Bigi:2000wn,Falk:2001hx,Falk:2004wg}.
These results also constrain many new physics
models~\cite{Burdman:2003rs,Petrov:2006nc,Golowich:2006gq,Golowich:2007ka,Golowich:2009ii}.
An observation of $\CP$ violation ($\CPV$) in the $\DzDzb$ system at the present experimental sensitivity
would provide possible evidence for physics beyond
the SM~\cite{Blaylock:1995ay,Isidori:2011qw,Hochberg:2011ru,Cheng:2012wr,Giudice:2012qq}.

One manifestation of $\DzDzb$ mixing is differing \Dz\ decay time
distributions for decays to different \CP eigenstates~\cite{Liu:1994ea}.
We present a measurement of charm mixing using the ratio of lifetimes obtained from the decays of neutral $D$ mesons to \CP-even
and \CP-mixed two-body final states. 
We also present a search for indirect \CP violation arising from a difference in \Dz and \Dzb partial decay widths to \CP-even
eigenstates. Recently the LHCb Collaboration has reported evidence for \CPV in the difference of the time-integrated \CP asymmetries
in $\Dz\to\KmKp$ and $\Dz\to\pimpip$ decays~\cite{Aaij:2011in}. 
This measurement is primarily sensitive to direct \CPV.
As explained in Appendix~\ref{app:mixing}, we are not sensitive to effects of direct \CP violation at the level of the result reported by LHCb,
and we therefore assume no direct \CPV in our baseline model.

We measure the effective \Dz lifetimes in three different
two-body final states: \DzRW, \KmKp, and \pimpip.
We make no distinction between the
Cabibbo-favored $\Dz \to \Kmpip$ and doubly Cabibbo-suppressed
$\Dz \to \Kppim$ modes; in other words, we analyze and describe them together.
Given the current experimental evidence indicating a small mixing rate,
the lifetime distribution for all two-body final states is exponential
to a good approximation.
Decays in the \DzRW mode are to a $\CP$-mixed final state, and are
assumed to be described by the average \Dz width $\Gamma$.
The singly Cabibbo-suppressed decays \Dz (\Dzb) to the \CP-even $\KK$ and $\pipi$  
final states are described by the partial decay rate $\Gamma^+$ ($\overline{\Gamma}^+$),
where $+$ indicates the \CP of the final state.
We present in  Appendix~\ref{app:mixing} a discussion of the
mixing formalism leading to the expressions that
are used to extract the mixing parameter $\yCP$ and the $\CPV$
parameter $\deltaY$,
\begin{eqnarray}
  \yCP &=&\frac{\Gamma^++\overline{\Gamma}^+}{2\Gamma} - 1 , \label{eq:yCPcalc2} \\
  \deltaY &=&  \frac{\Gamma^+-\overline{\Gamma}^+}{2\Gamma},
  \label{eq:deltaYcalc2}
\end{eqnarray} 
 from the experimentally measured $\CP$-mixed and $\CP$-even lifetimes.
This definition of \deltaY is opposite in sign to that in our previous measurement~\cite{Aubert:2007en} and is now consistent with that used by the Heavy Flavor Averaging Group~\cite{Asner:2010qj}.

We use \Dz mesons from {\sl tagged} $\Dstp \to \Dz \pip$
decays~\cite{CC:2008}, as well as {\sl untagged} decays where
no $\Dstp$ parent is found.
The charge of the $D^{*\pm}$ is used to split the \KmKp and \pimpip
samples into those originating from \Dz and from \Dzb
mesons in order to measure the \CP-violating parameter $\deltaY$.
The requirement of a \Dstp\ parent strongly suppresses backgrounds; hence untagged decays are reconstructed only in \DzRW and \KmKp
because of the relatively poor signal-to-background ratio in the untagged \pimpip final state.
In summary we study seven modes, two untagged and five tagged.

In addition to the increased integrated luminosity of the new dataset compared to that used in our earlier results~\cite{Aubert:2007en,Aubert:2009ai},
 this analysis benefits from improved particle identification and charged-particle track reconstruction, and a more inclusive and optimized event selection.
We implement an improved background model, and we simultaneously fit both the tagged and untagged datasets.

\section{Event Reconstruction and Selection}

We use $468 \invfb$ of $\epem$ colliding-beam data recorded at, and slightly
below, the $\FourS$ resonance ($\epem$ center-of-mass [CM] energy $\sqrt{s} \sim 10.6\gev$)
with the \babar\ detector~\cite{Aubert:2001tu}
at the SLAC National Accelerator Laboratory \hbox{PEP-II} asymmetric-energy $B$ Factory.
To avoid potential bias, we finalize our data selection criteria,
as well as the procedures for fitting, extracting statistical limits,
and determining systematic uncertainties, prior to examining the results.

We reconstruct charged tracks and vertices with a 5-layer, double-sided silicon vertex tracker (SVT)
and a 40-layer drift chamber (DCH). We select \Dz candidates by pairing oppositely charged tracks,
requiring each track to satisfy particle identification criteria based on specific ionization energy loss (\dedx)
from the SVT and DCH, and Cherenkov angle measurements from a ring-imaging Cherenkov detector (DIRC).
We then refit the \Dz daughter tracks, requiring them to originate from a common vertex.
To reduce contributions from \Dz's
produced via $B$-meson decay to a negligible level,
we require each \Dz to have momentum in the CM frame $p_{\rm CM} > 2.5\gevc$.

For tagged decays, we reconstruct \Dstarp candidates by combining a \Dz candidate with a
slow pion track $\pisoftp$, requiring them to originate
from a common vertex constrained to the \epem interaction region.
We require the $\pisoftp$ momentum
to be greater than $0.1\gevc$ in the laboratory frame and less than
$0.45\gevc$ in the CM frame.
We reject a positron that fakes a $\pisoftp$ candidate
by using \dedx information and veto
any $\pisoftp$ candidate that may have originated from a reconstructed
photon conversion or \piz Dalitz decay.
The distribution of the difference $\dm$ between the reconstructed $D^{*+}$ and
$\Dz$ masses peaks near $\dm \sim 0.1455\gevcc$.  Backgrounds are
suppressed by retaining only  tagged candidates in the range $0.1447 < \dm < 0.1463\gevcc$.

To determine the proper time $\t$ and its error $\terr$ for each \Dz candidate,
we perform a combined fit to the \Dz production and decay vertices. We constrain the 
production point to be within the $\epem$ interaction region, which we
determine using Bhabha and di-muon events from triggers close in time to any given signal candidate event.
We retain only candidates with a \chisq-based probability
for the fit $\Pchisq > 0.1\%$, and with $-2 < \t < 4 \ps$
and $\terr < 0.5 \ps$.
For tagged decays, this fit  does not
incorporate any $\pisoftp$ information in order
to ensure that the lifetime resolution models for tagged and untagged
signal decays are very similar.
The most probable value of \terr for signal events
is $\sim 40\%$ of the nominal $\Dz$ lifetime~\cite{Nakamura:2010zzi}.

If an event contains a tagged \Dz decay, we exclude all untagged \Dz candidates from that event in the final sample.
For a given final state, when multiple \Dz (for the untagged modes) or \Dstp (for the tagged modes) 
candidates in an event share one or more tracks, we retain only the candidate with the highest
\Pchisq. The fraction of events with multiple $\Dz$ candidates with overlapping
daughter tracks is $\ll 1\%$ for all final states.

\section{Invariant Mass Fits}

We characterize the \Dz invariant mass ($M$) distribution for each of the seven modes with an extended unbinned
maximum likelihood fit to \Dz and \Dzb samples.
We allow the parameters governing the shapes of the probability density functions (PDFs),
 as well as the expected signal and background candidate yields, to vary in the fits. 
For the tagged \CP-even modes we fit the \Dz and \Dzb samples simultaneously, sharing all parameters except for the 
expected signal and background candidate yields.
 
We fit the tagged $\pipi$ invariant mass distribution in the fit range $1.82<M_{\pi\pi}<1.93 \gevcc$
using a sum of two Gaussians with independent means and widths for the signal PDF,
along with a first-order Chebychev polynomial for the total background.

The fit model for the tagged $\KK$ invariant mass distribution is similar to $\pipi$,
except that the fit range is $1.82<M_{KK}<1.91 \gevcc$,
and the signal PDF is the sum of two independent Gaussians
and a modified Gaussian with a power-law tail~\cite{Oreglia:1980cs}, which aids in better modeling of
the lower tail of the distribution.

The signal PDF for the untagged $\KK$ mode and for both tagged and untagged $\DzRW$ modes is a sum of three independent Gaussians;
the background is modeled using a second-order Chebychev polynomial.
The mass fit range is $1.82<M_{KK}<1.91 \gevcc$ for the untagged \KK mode, $1.81<M_{K\pi}<1.92 \gevcc$ for the untagged \DzRW mode, and $1.80<M_{K\pi}<1.93 \gevcc$ for the tagged \DzRW mode.
In these modes, we do not distinguish $\Dz$ from $\Dzb$ candidates, and therefore determine 
only the total signal and total background yields, in addition to the signal and background shape parameters.

\begin{figure}[!ht]
  \centerline{
  \includegraphics[width=0.5\linewidth]{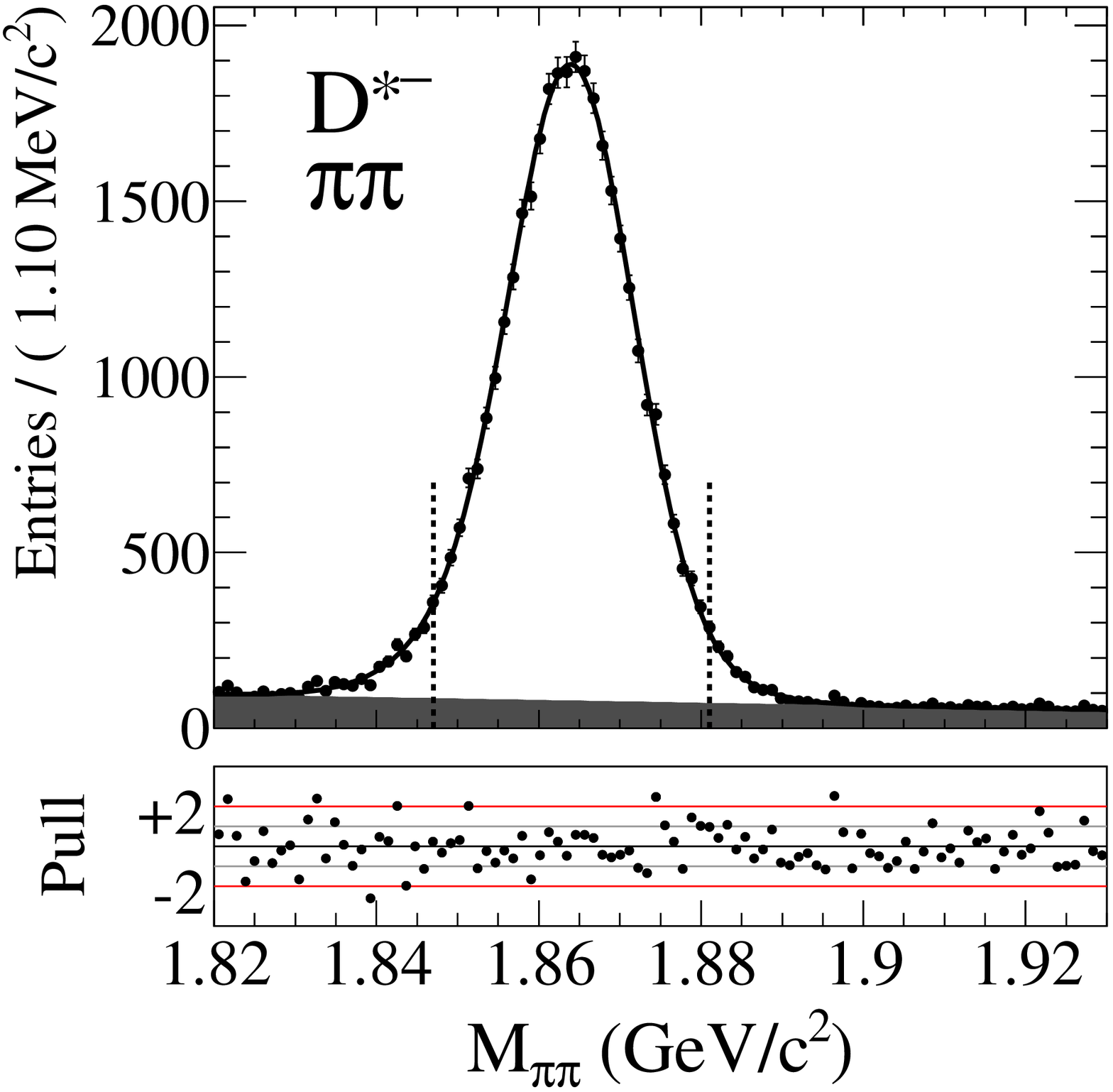}\hfill
  \includegraphics[width=0.5\linewidth]{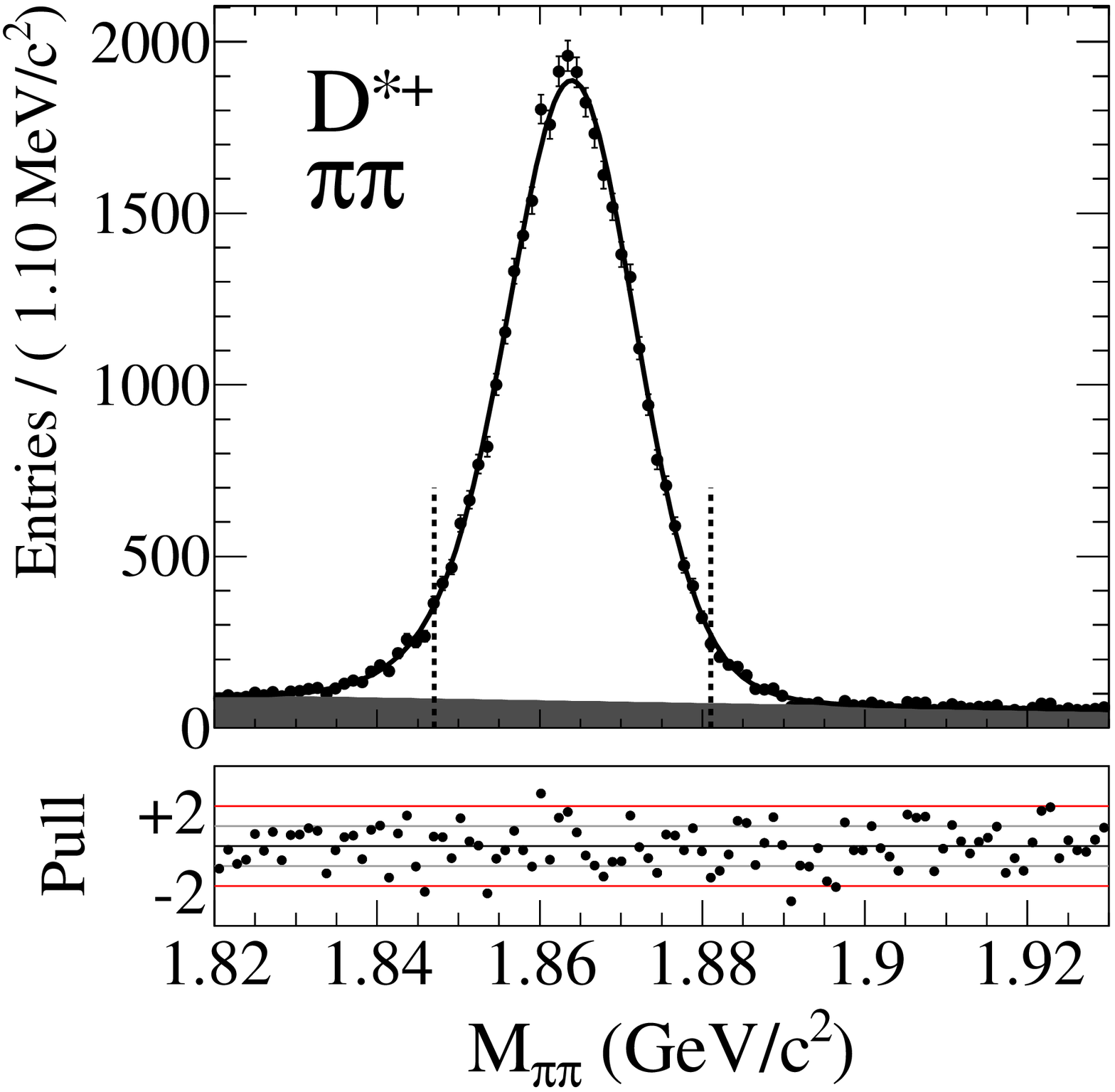}
}
  \centerline{
  \includegraphics[width=0.5\linewidth]{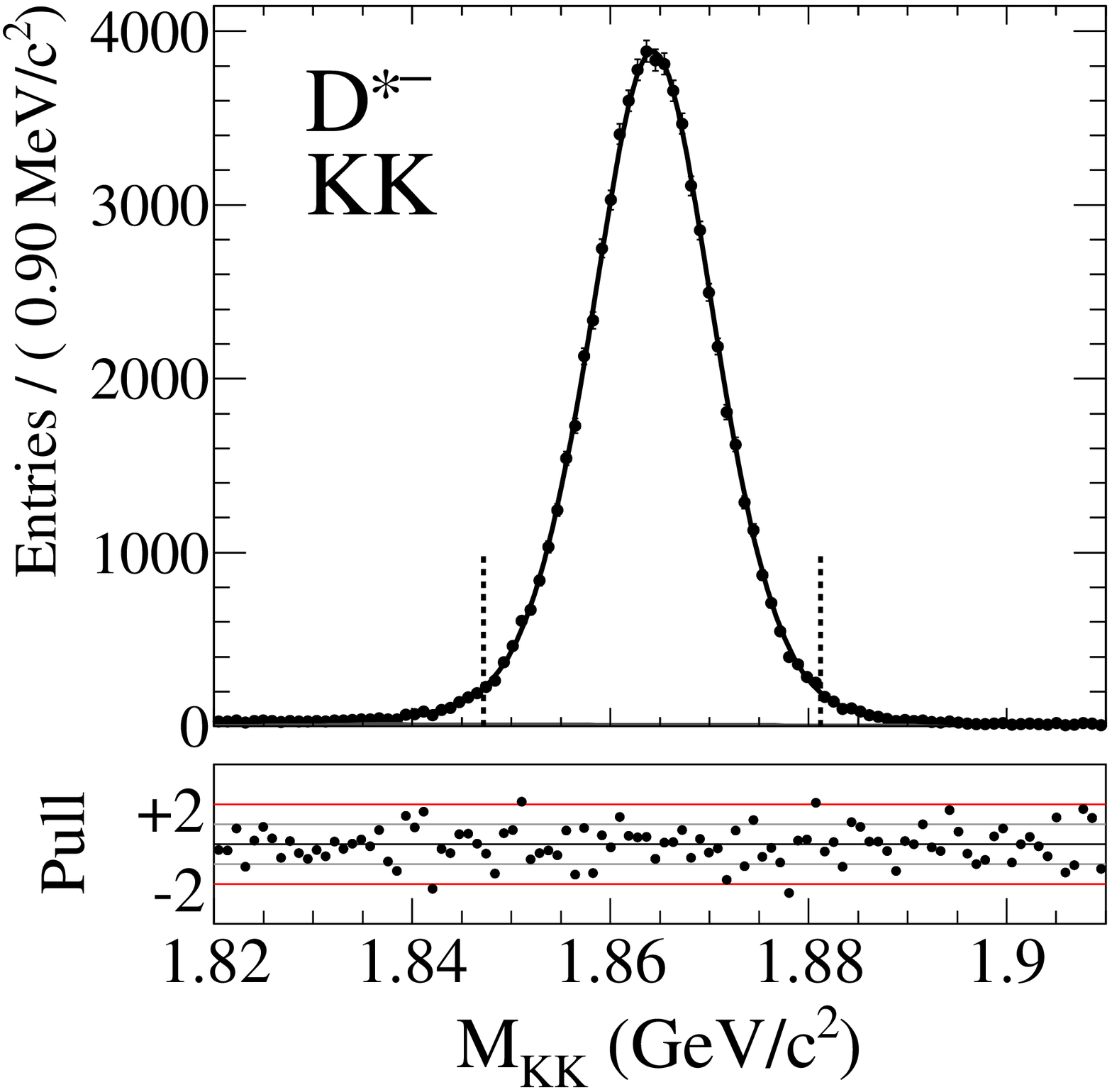}\hfill
  \includegraphics[width=0.5\linewidth]{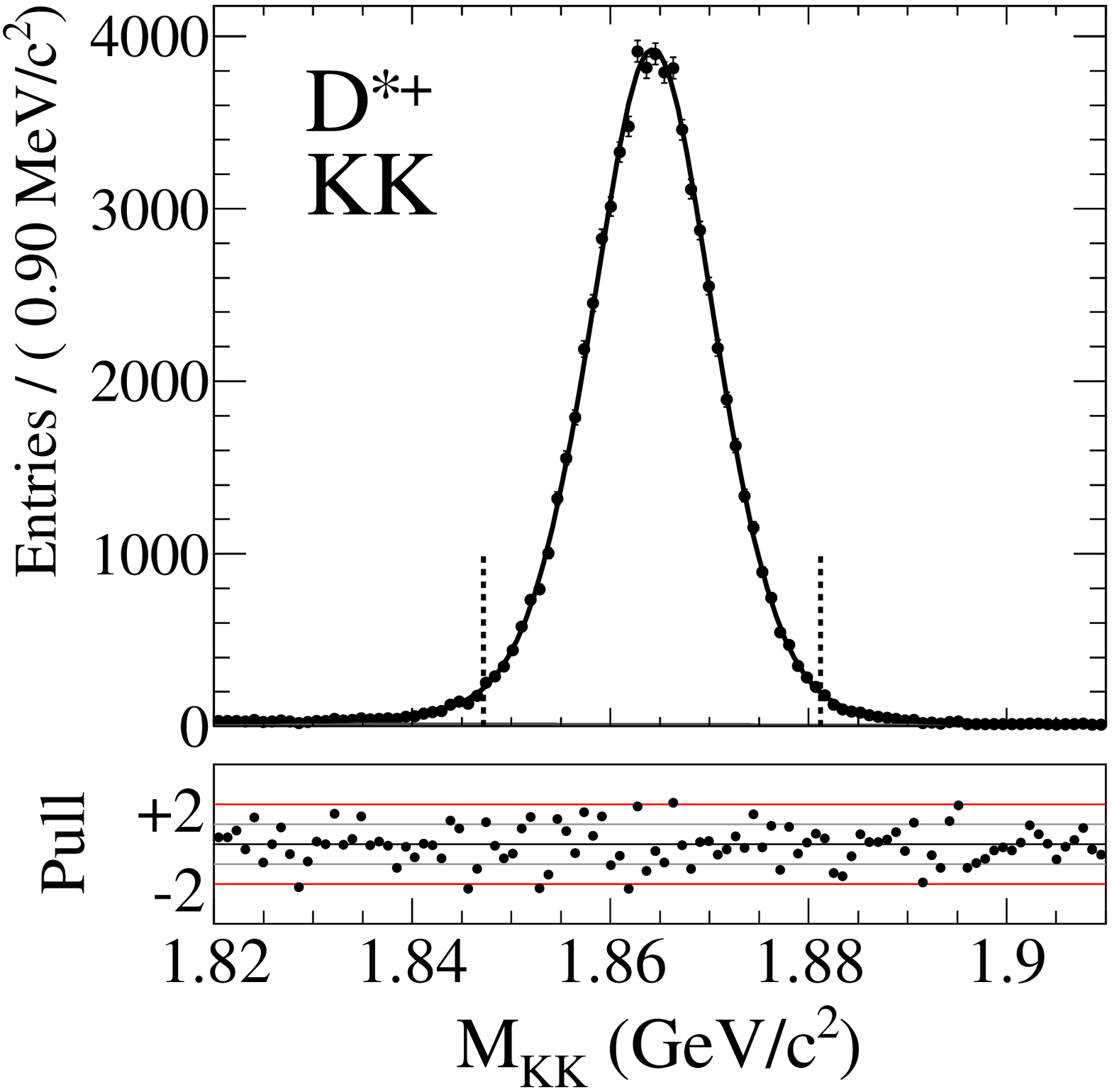}
}
  \centerline{
  \includegraphics[width=0.5\linewidth]{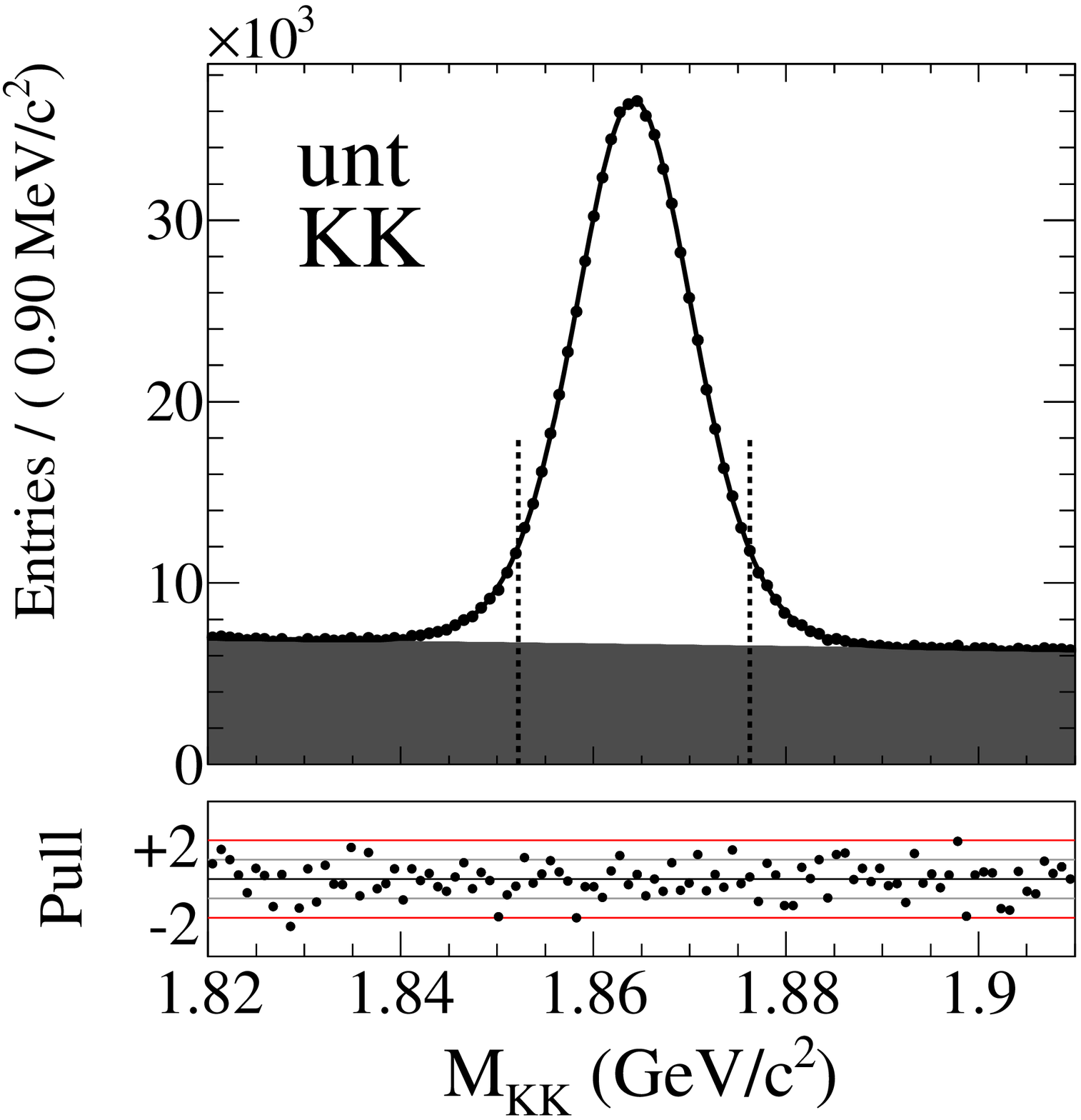}\hfill
 \includegraphics[width=0.5\linewidth]{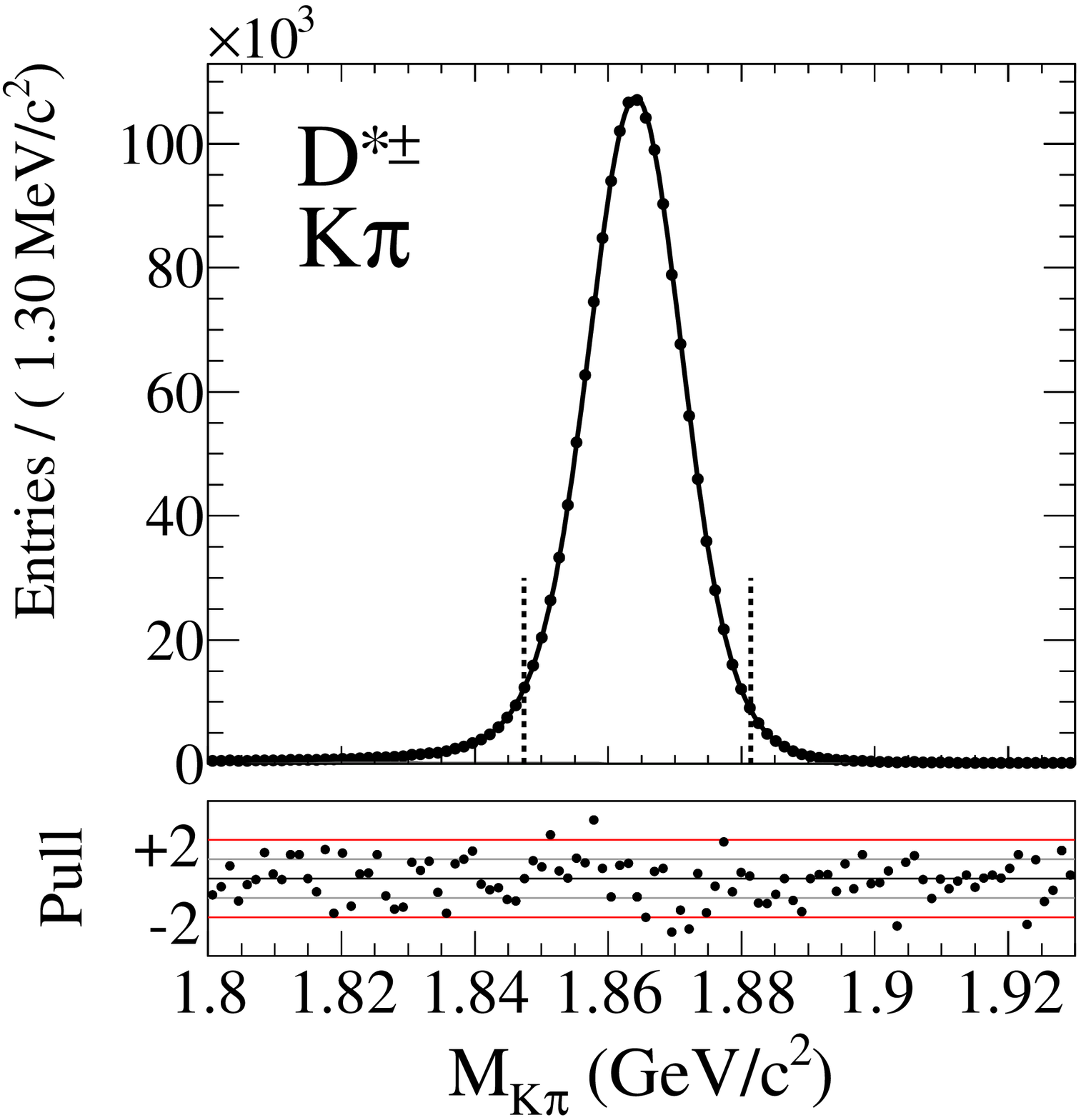}
}
  \centerline{
 \includegraphics[width=0.5\linewidth]{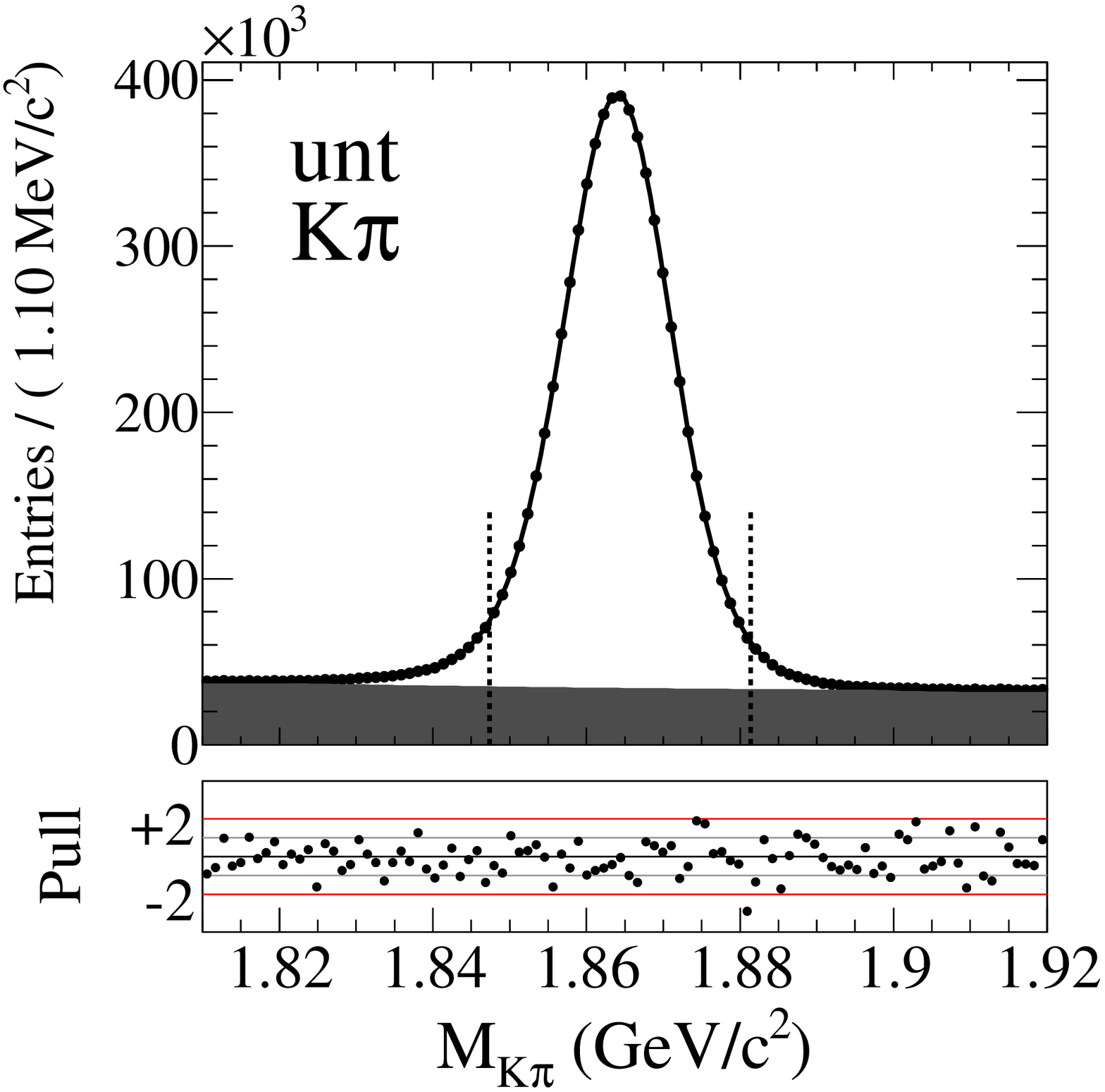}\hfill
}
  \caption{The reconstructed two-body invariant mass distributions
    for the seven modes. The vertical lines show the \emph{lifetime-fit mass
    region}, defined in Sec.~\ref{sec:sigBox}. The shaded regions are the background contributions.
    The normalized Poisson pulls for each fit are shown under each plot; ``unt'' refers to the untagged datasets. }
\label{fig:MassPlots}
\end{figure}

The reconstructed \Dz invariant mass distributions and the fit results are shown in Fig.~\ref{fig:MassPlots},
together with a plot of the corresponding normalized Poisson pulls~\cite{Baker:1983tu}. 

\section{Signal and Sideband Regions}
\label{sec:sigBox}

For the lifetime fit, we determine the regions in two-body invariant
mass that maximize signal significance, minimize systematic effects due
to backgrounds, and minimize the effect of the correlation between \Dz invariant mass and proper time.
We refer to these regions as the \emph{lifetime-fit mass regions}.
Based on these studies, the optimal
lifetime-fit mass region is $34\mevcc$ wide for all tagged modes
and untagged $\DzRW$ events, $1.847<M<1.881 \gevcc$.
Because of the smaller signal-to-background ratio for the
untagged $\KK$ events, the lifetime-fit mass region for this
mode is only $24\mevcc$ in width, $1.852<M<1.876 \gevcc$.
For the tagged modes, a mass difference
sideband $0.151 < \dm < 0.159 \gevcc$ is used, along with a low (high)
invariant mass sideband, $1.819\ (1.890)<M<1.839\ (1.910) \gevcc$.
The low (high) mass sideband used for the untagged modes, $1.810\ (1.899)<M<1.830\ (1.919) \gevcc$, is displaced from the tagged sideband
in order to reduce the signal component there.
The signal purities in the lifetime-fit mass regions range from $\sim 75\%$ for the
untagged $\KK$ sample to $\sim 99.8\%$ for the tagged $\DzRW$ events.

We classify \Dz candidate decays in the lifetime-fit mass region as follows: 
 \Dz signal decays; misreconstructed-charm decays, {\it i.e.}, those in which the candidate-\Dz daughter tracks are
decay products of a non-signal weak charm decay; and random combinatorial background.
Table~\ref{tab:charm} gives the composition of the
misreconstructed-charm backgrounds expected from simulated events~\cite{simulation} in each final state.

\begin{table}[!h]
\caption{Expected composition (in \%) of the mis\-re\-con\-structed-charm backgrounds.
Only misreconstructed-charm background
channels that have $>1\%$ contribution in at least one signal mode are listed.
For the tagged modes, the yields are the sum of the separate $\Dz$ and $\Dzb$ tags.}
\begin{center}
\begin{tabular}{llllll}
\hline \hline
\multirow{2}{*}{Mode} & \multicolumn{3}{c}{Tagged Modes} & \multicolumn{2}{c}{Untagged Modes}     \\ 
                      & \pimpip        & \KmKp          & \DzRW          &  \KmKp         &  \DzRW  \\ 
\hline 
$\Dz\to X\ell\nu$     & 15.4           & 10.3           & 29.9           & \phantom{0}7.2 & $\, \le 2$  \\
$\Dz\to\Km\pip$       & 80.8           & 14.9           & 57.1           & \phantom{0}8.8 & 35.8     \\
$\Dz\to\piz\pip\Km$   & \phantom{0}1.1 & 70.3           & \phantom{0}1.7 & 63.3           & \phantom{0}6.9      \\
$\Dp\to\pip\pip\Km$   & $\, \le 1$     & \phantom{0}2.9 & $\, \le 1$     & 11.8           & $\, \le 2$  \\
$\Dz\to\Kp\Km$        & $\, \le 1$     & $\, \le 1$     & \phantom{0}1.3 & $\, \le 1$     & \phantom{0}3.5      \\
$\Dz\to\pip\pim$      & \phantom{0}1.8 & $\, \le 1$     & \phantom{0}2.2 & $\, \le 1$     & \phantom{0}3.1      \\
$\Dz\to\pip\pim\piz$  & $\,\le 1$      & $\, \le 1$     & \phantom{0}7.0 & $\, \le 1$     & 17.3     \\
$\Lambda$ decays      & $\,\le 1$      & $\, \le 1$     & $\, \le 1$     & \phantom{0}4.9 & \phantom{0}2.6      \\
\hline\hline
\end{tabular}
\label{tab:charm}
\end{center}
\end{table}

\section{lifetime fit}

The lifetimes are determined from an extended unbinned maximum likelihood fit
to $t$ and $\terr$ for candidates in the lifetime-fit mass region. All modes are 
fit simultaneously using shared signal resolution function parameters.
The signal, misreconstructed-charm and combinatorial components are described by their own set of PDFs,
which in the tagged modes can also depend on the charm flavor.

The  lifetime PDF for signal is an exponential function convolved with a resolution function,
which is  the sum of three Gaussian functions whose widths are
proportional to $\terr$.
The explicit form of the signal lifetime PDF is
\begin{eqnarray} 
\label{eq:sharedsignaltimeDistribution}
{\cal R}_{F,L}^T(t,\terr) &=& f_{t1}{\cal D}(t,\terr;\Sprime \SX s_1,\toff,\tau_L) \\  \nonumber
&+& ( 1 - f_{t1})\Big[ f_{t2}{\cal D}(t,\terr;\Sprime \SX s_2,\toff,\tau_L) \\  \nonumber
&+& ( 1 - f_{t2}){\cal D}(t,\terr;\Sprime \SX s_3,\toff,\tau_L)\Big] ,
\end{eqnarray} 
 where $f_{ti}$ (with $i=1,2$)
parameterizes the contribution of each individual Gaussian,
$s_i$ (with $i=1,2,3$) is a scaling factor associated with each Gaussian, and
$\toff$ is an offset of the mean of the resolution function. The function ${\cal D}(t,\terr;s,\toff,\tau)$
is given by
\begin{equation}
\label{eq:test}
\begin{array}{l}
{\cal D}(t,\terr;s,\toff,\tau) =\\[5pt] 
\quad
C_{\terr}{\displaystyle\int} \exp(-\ttrue/\tau)
                               \exp\left( -\frac{(t-\ttrue+\toff)^2}{2(s\cdot\terr)^2}\right)\, d\ttrue,
\end{array}
\end{equation}
\noindent where the normalization coefficient $C_{\terr}$ is chosen such that
\begin{eqnarray} 
  \int {\cal D}(t,\terr;s,\toff,\tau)\, dt=1\quad\hbox{for each \terr.}
\end{eqnarray} 
With this definition, the product $\RterrSig(\terr)\cdot{\cal D}(t,\terr;s,\toff,\tau)$
is a properly normalized two-dimensional conditional PDF, where \RterrSig(\terr) is a
PDF characterizing the $\terr$ distribution, described below.
To account for small differences in the resolution function for the different final states
we introduce additional mode-dependent scale factors $\SX$, $F = K\pi ,\ KK ,\ \pi\pi$.
We also allow for differences between the resolution functions for tagged and untagged modes
by means of scale factors \Sprime, $T=$ tag (tagged) or unt (untagged).
We fix $\SKpi$ and $\Sunt$ to 1.

The three lifetime parameters are $\tau_L = \{\tauhhp,\tauhhm, \tauKpi\}$, where $\tauKpi$ is extracted from the tagged and untagged
$\DzRW$ modes, while $\tauhhp$ and $\tauhhm$ are extracted from the tagged and untagged \CP-even modes.
Approximately 0.4\% of the tagged \CP-even samples contain correctly reconstructed \Dz candidates combined
with an unrelated \pisoftp; this fraction has been estimated from simulated 
events and verified in data by an earlier  \babar\ analysis~\cite{Aubert:2007wf}.
 These candidates have the same resolution and lifetime behavior as those from correctly reconstructed \Dstp\ decays,
but about half of them will be tagged as the wrong flavor. Therefore, the tagged \CP-even \Dz proper-time distributions
are modeled as the weighted sum of PDFs for correctly tagged and untagged candidates, characterized by the lifetime parameters
\tauhhp and \tauhhm, respectively, and a mistag fraction $f_{\rm tag} = 0.2\%$.
The tagged \CP-even \Dzb proper-time distributions are modeled in a similar fashion, where now the correctly tagged and mistagged PDFs are characterized by the lifetime parameters \tauhhm and \tauhhp, respectively.
The untagged \KmKp proper-time distribution is modeled as a weighted sum of two PDFs characterized by the lifetime parameters \tauhhp and \tauhhm,
 respectively, and a weighting fraction $f_{\Dz} = 0.5$.
These parameterizations assume no direct \CPV, and allow for \CPV in the interference between  decays with and without mixing characterized by a mode-independent weak phase $\phi$.
Both $f_{\rm tag}$ and $f_{\Dz}$ are varied as part of the systematic error estimate for \yCP and \deltaY.
All five tagged and two untagged signal lifetime PDFs are explicitly given in Appendix~\ref{app:pdf}.

The $\terr$ PDF for signal candidates is obtained directly
from data by subtracting the sum of the background $\terr$ distributions from
that of all candidates in the lifetime-fit mass region. 
These 1-d $\terr$ distributions are used to model the $\RterrSig(\terr)$
PDF discussed previously.

We determine the $t$ versus $\terr$ misreconstructed-charm signal-like PDF shape parameters and yields
by fitting simulated events in the lifetime-fit mass region and then fix these parameters in the lifetime fit to data.
We vary the lifetimes and yields as part of the study of systematic effects.

The largest background in the lifetime-fit mass region is due to random combinations of tracks.
The PDF describing the two-dimensional combinatorial background in $t$ and $\terr$ in the
lifetime-fit mass region is characterized as a weighted average of the 2-d PDFs extracted
from the mass sideband regions.
The weights for the low and high sidebands are obtained from simulated events.
The $(t,\terr)$ combinatorial PDF in each sideband and for each mode, except for the untagged \KmKp
mode, is extracted as a 2-d histogram from the sideband samples. From
these histograms we subtract the contribution of signal and
misreconstructed-charm backgrounds, each of which is estimated from
simulated events, to obtain the final combinatorial PDF in each sideband.
For the untagged $\KK$ mode, a similar procedure is used but, instead of
histograms, analytic signal-like PDFs are used.
For the background PDFs the offsets and the lifetimes are allowed to be different
for each Gaussian.
 The signal and misreconstructed-charm PDF parameters are extracted by fitting simulated 
events and then fixed, along with the expected candidate yields, in the fit that extracts the combinatorial PDFs in each sideband.

For the untagged $\KmKp$ mode both the expected signal and combinatorial yields
are free parameters in the lifetime fit.
The expected combinatorial background yields in the other modes are
determined by integrating the total background PDF extracted from the mass fit in the lifetime-fit
mass region, and then subtracting the expected misreconstructed-charm background yields, which are determined from samples of simulated events. 
A small bias on these fit yields is observed in fits to
simulated events. To correct for this, we scale the data yields based on
the simulated-event fits and vary the mode-dependent scale factors as a systematic uncertainty.
Table~\ref{tab:yields} gives the event class yields plus uncertainties obtained from the lifetime fit
and indicates the yields that are fixed.

\begin{table}[!h]
\caption{Signal and background yields in the lifetime-fit mass region. Yields with
uncertainties are those obtained directly from the lifetime fit to data. For the tagged modes,
the yields are the sum of the separate $\Dz$ and $\Dzb$ tags.}
\begin{center}
\begin{tabular}{lrrrrr}
\hline \hline
            & \multicolumn{3}{c}{Tagged Modes} & \multicolumn{2}{c}{Untagged Modes}     \\ 
            & \pimpip &  \KmKp    &  \DzRW       &  \KmKp         &  \DzRW     \\ 
 \hline
Signal      & $65\,430$ & $136\,870$  & $1\,487\,000$ & $496\,200$  & $5\,825\,300$ \\
            & $\pm 260$ & $ \pm 370$  & $\pm 1200 $   & $\pm 1200$  & $ \pm 2600 $ \\
Comb. Bkgd. & 3760      &        653  &  2849         & $165\,000$  & $1\,044\,552$            \\
            &           &             &               & $\pm 1000 $ & \\
Charm Bkgd. &   97      &        309  &   642         &   5477      & 4645               \\
\hline \hline 
\end{tabular}
\label{tab:yields}
\end{center}
\end{table}

\begin{figure}[!ht]
  \centerline{
    \includegraphics[width=0.5\linewidth]{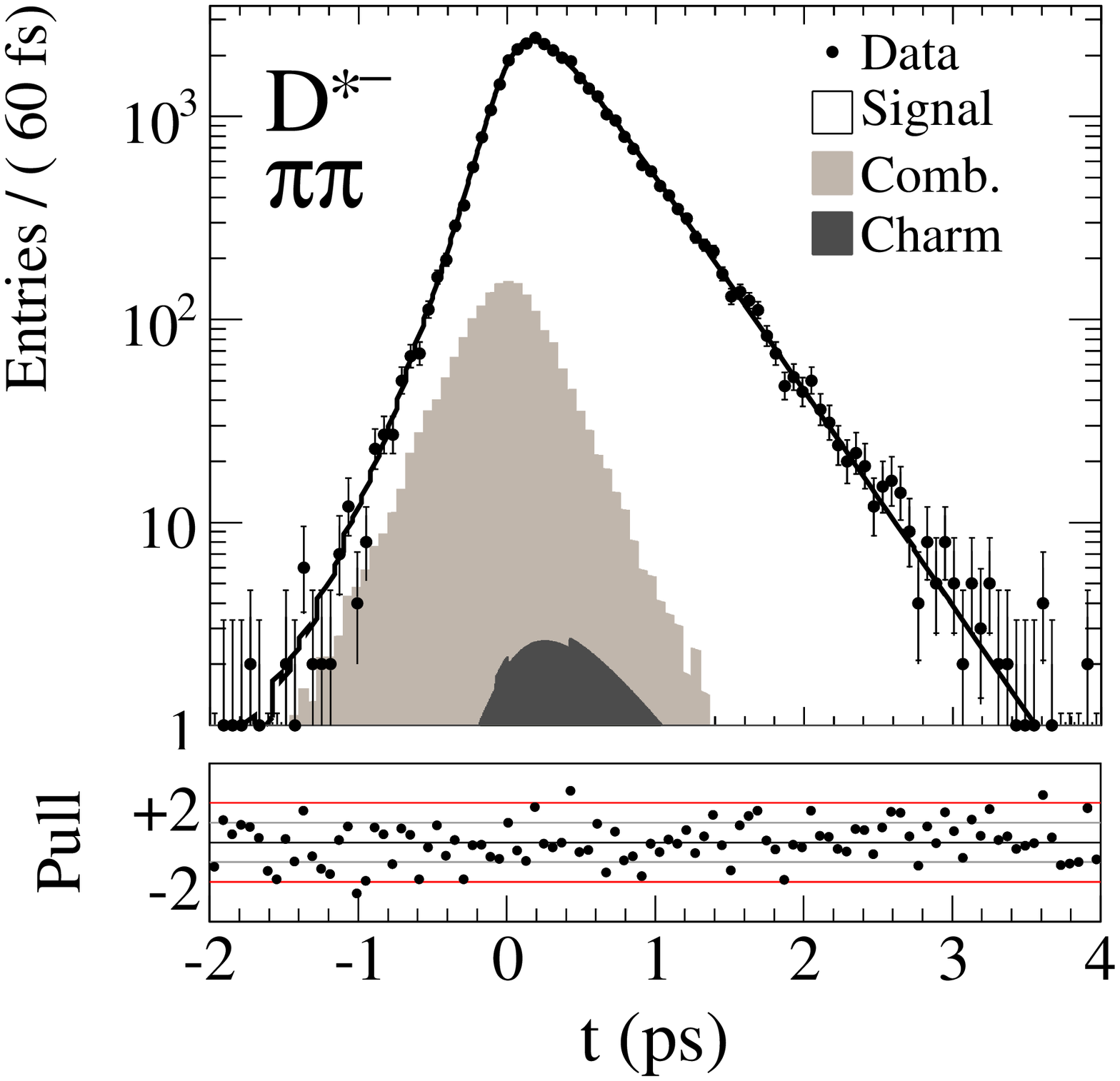}\hfill
    \includegraphics[width=0.5\linewidth]{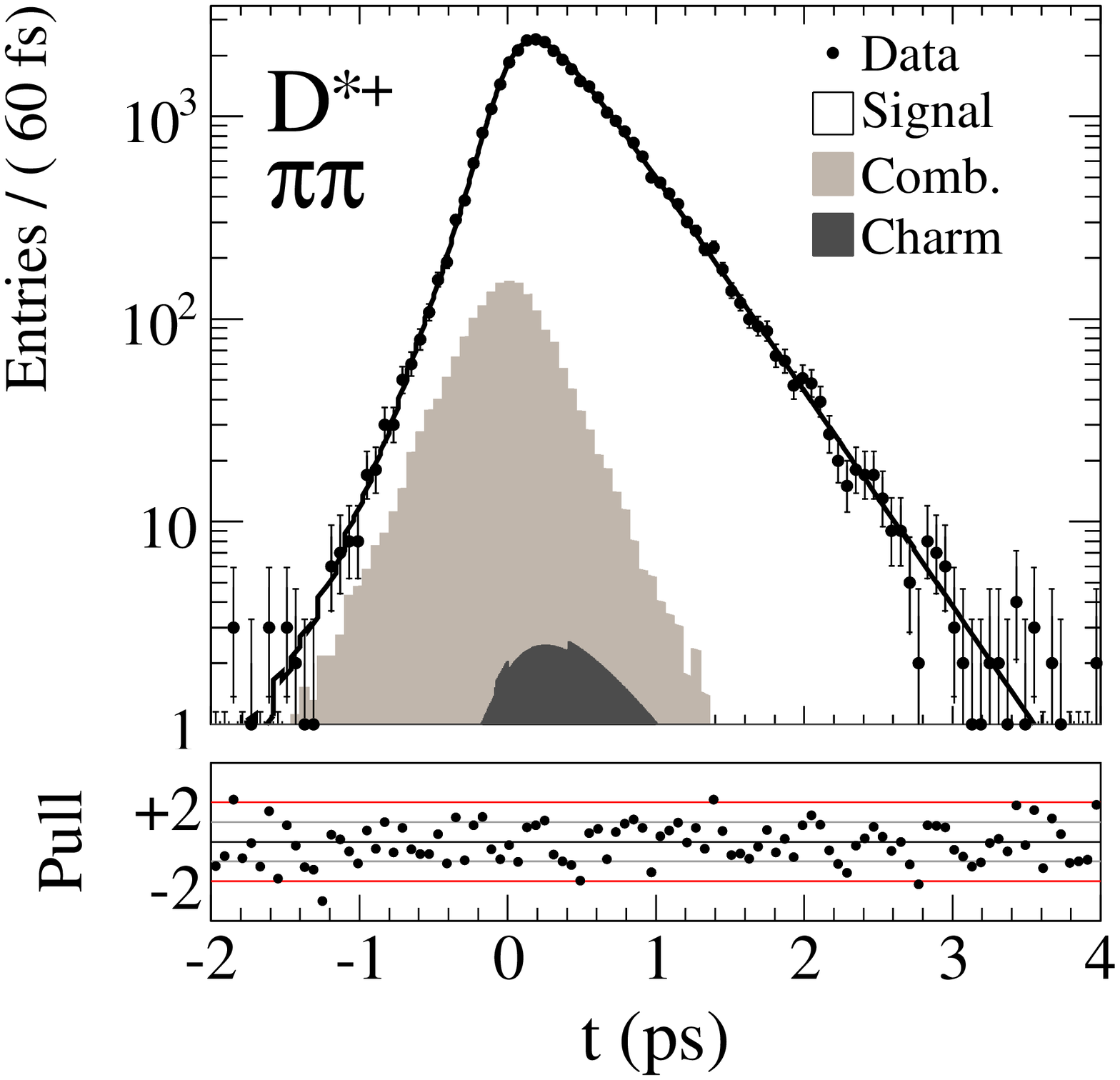}
  }
  \centerline{
    \includegraphics[width=0.5\linewidth]{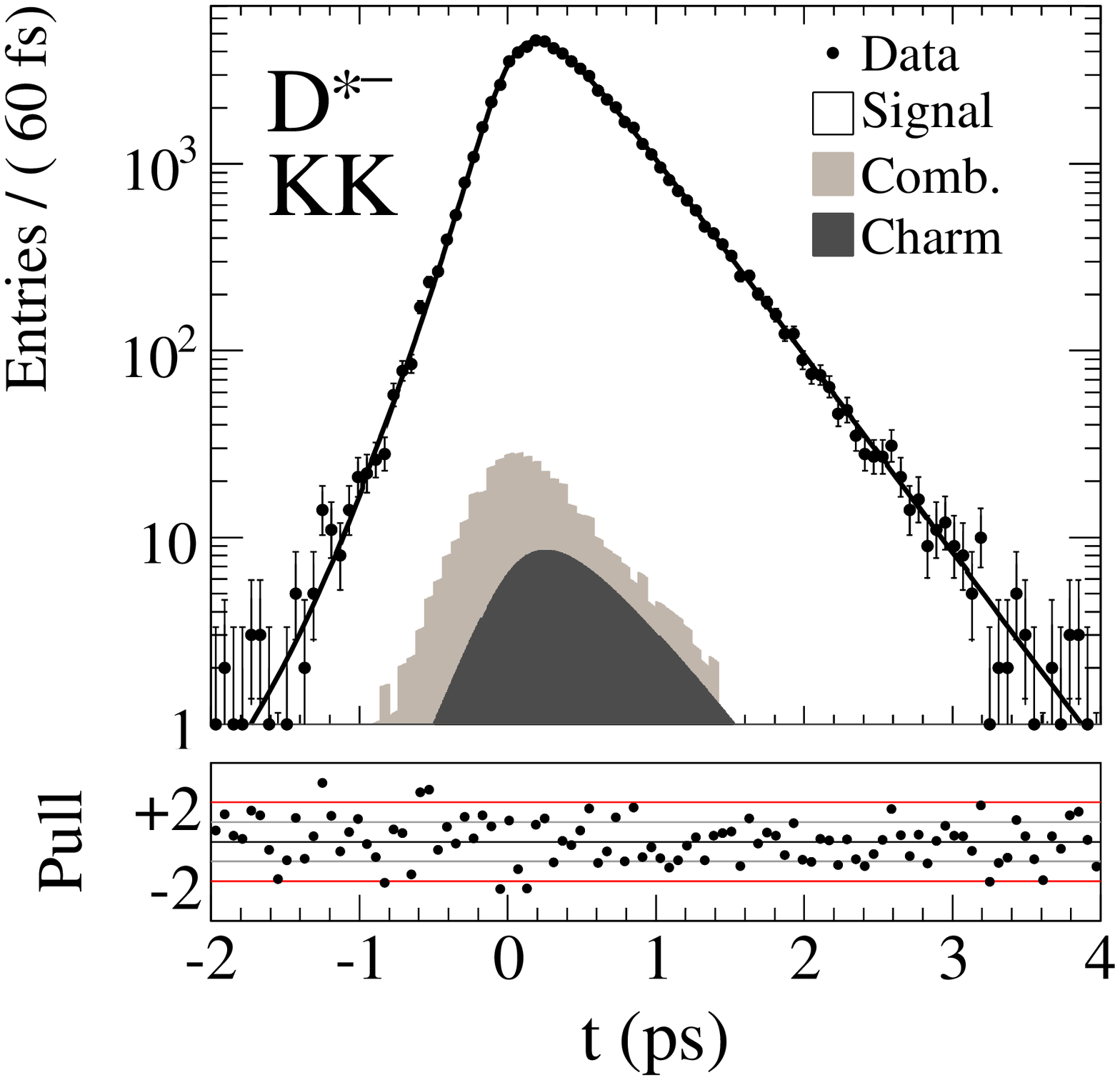}\hfill
    \includegraphics[width=0.5\linewidth]{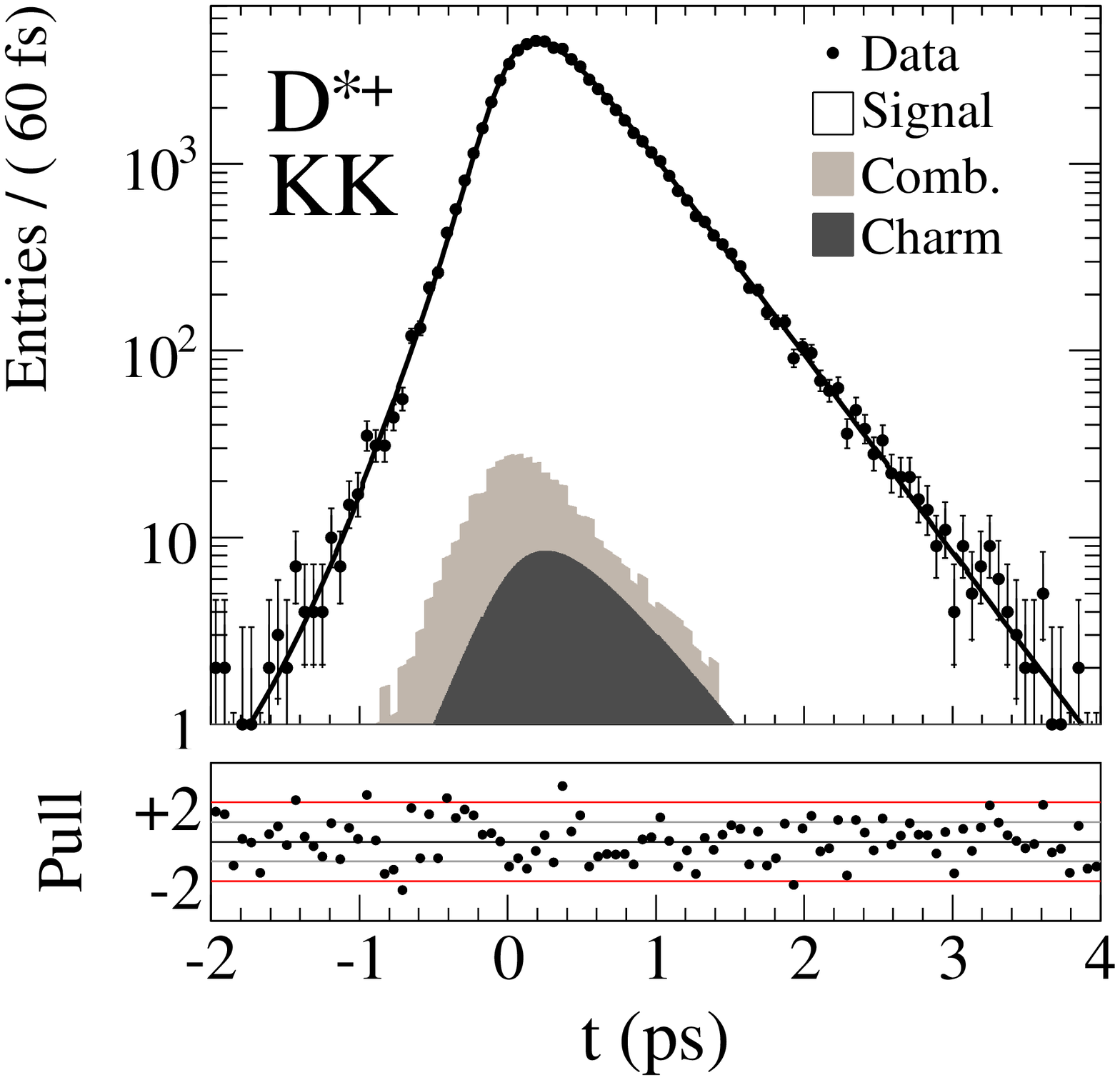}
  }
  \centerline{
    \includegraphics[width=0.5\linewidth]{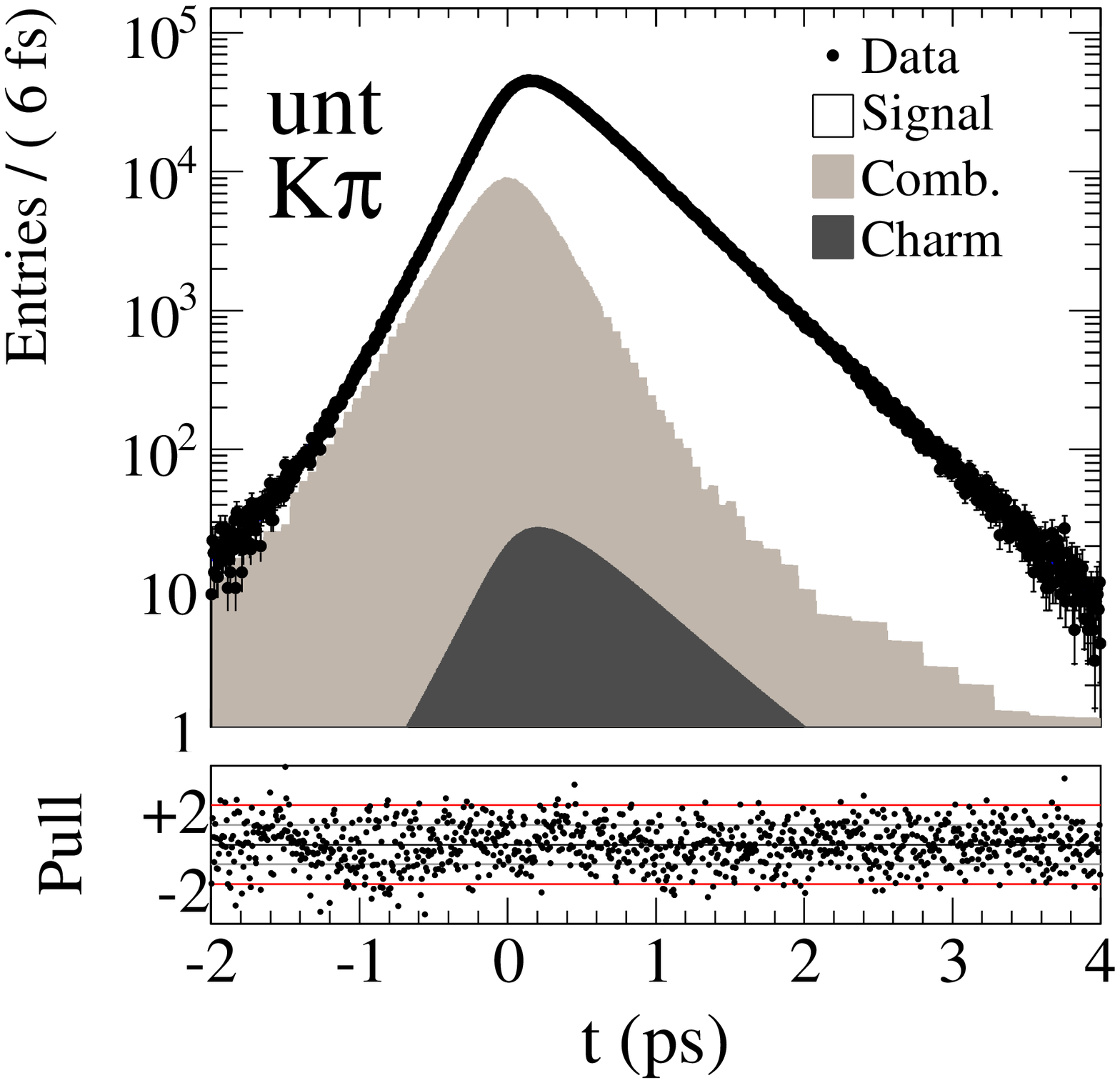}\hfill
    \includegraphics[width=0.5\linewidth]{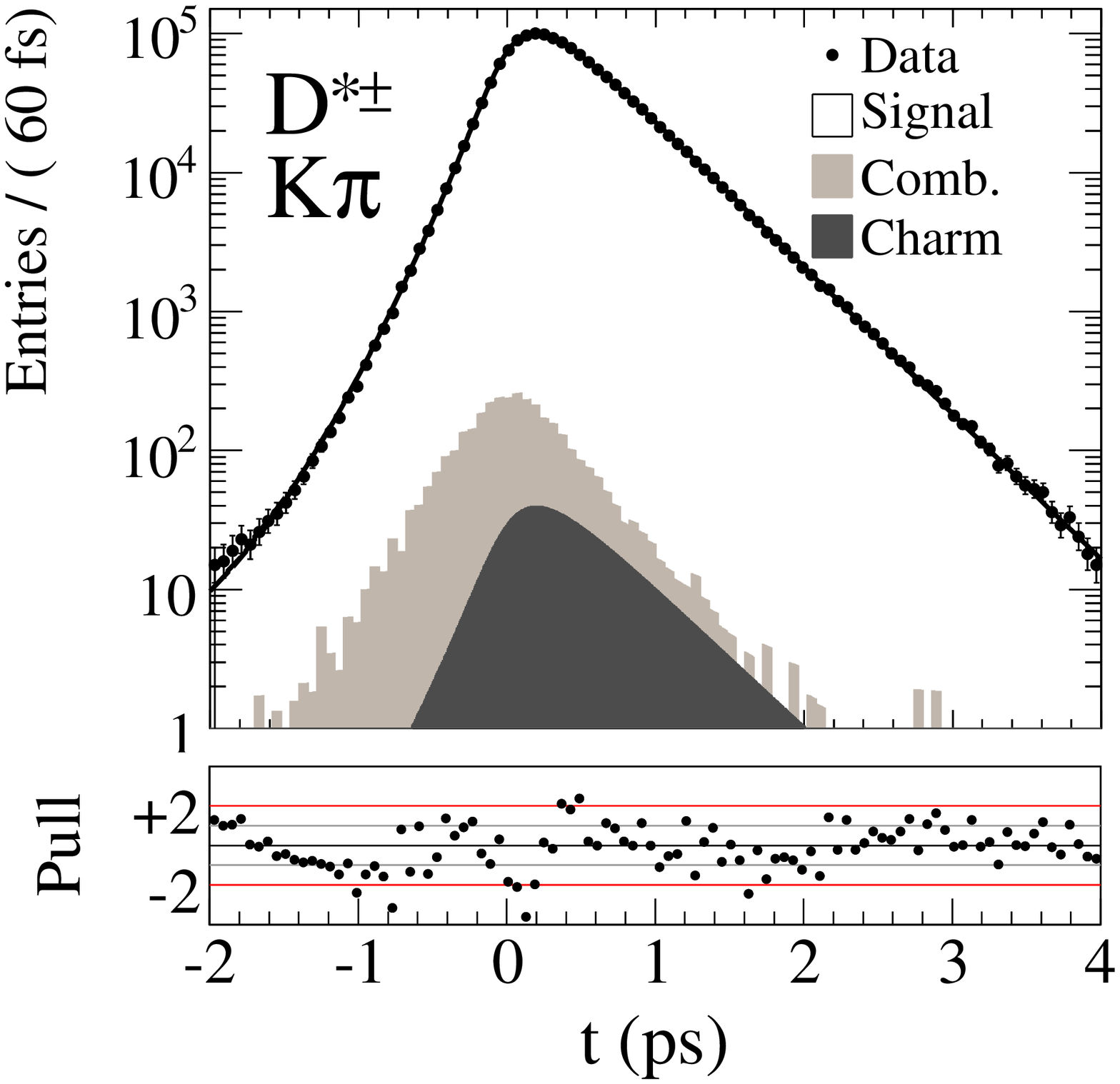}
  }
  \centerline{
    \includegraphics[width=0.5\linewidth]{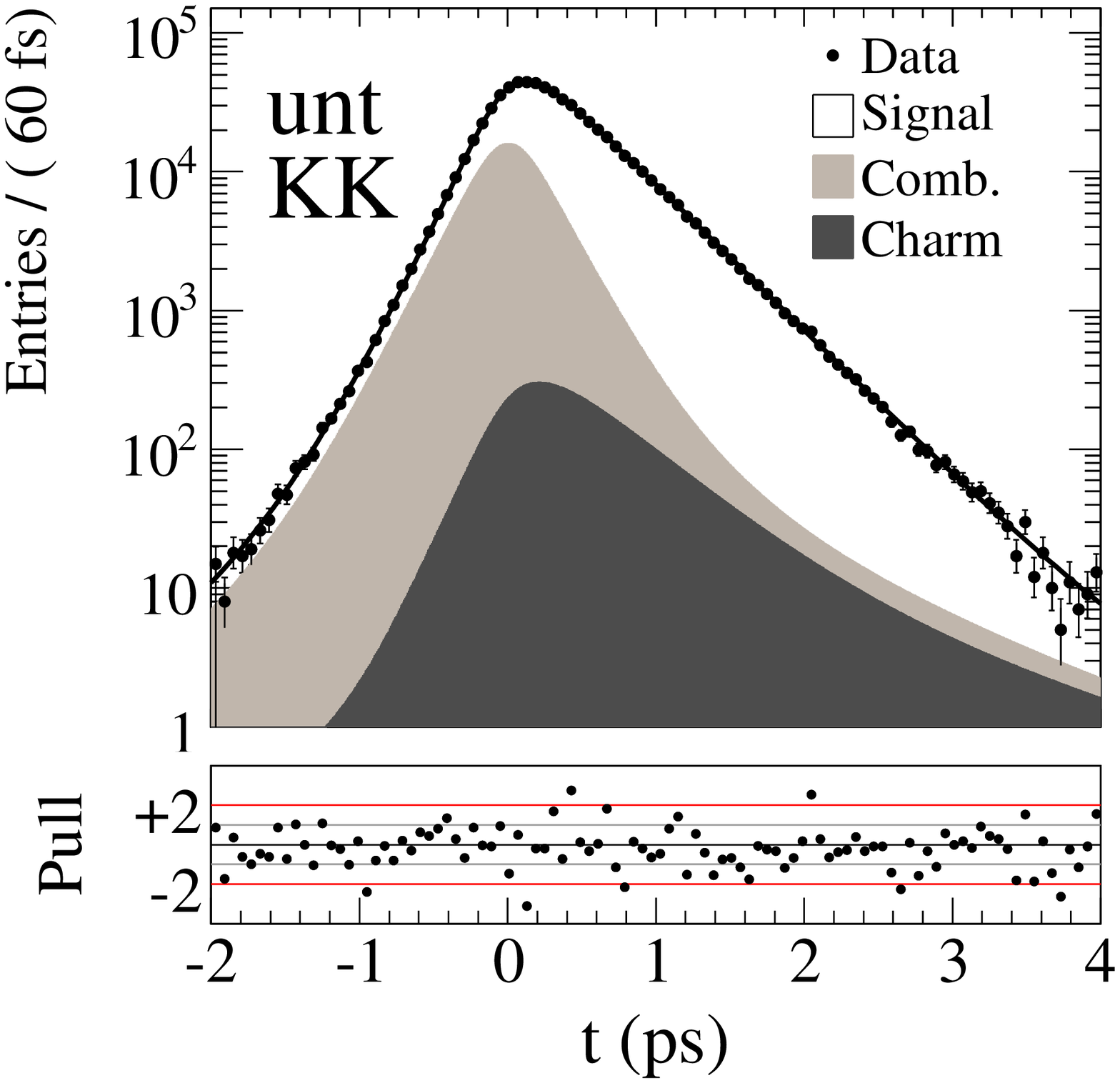}\hfill
    \includegraphics[width=0.5\linewidth]{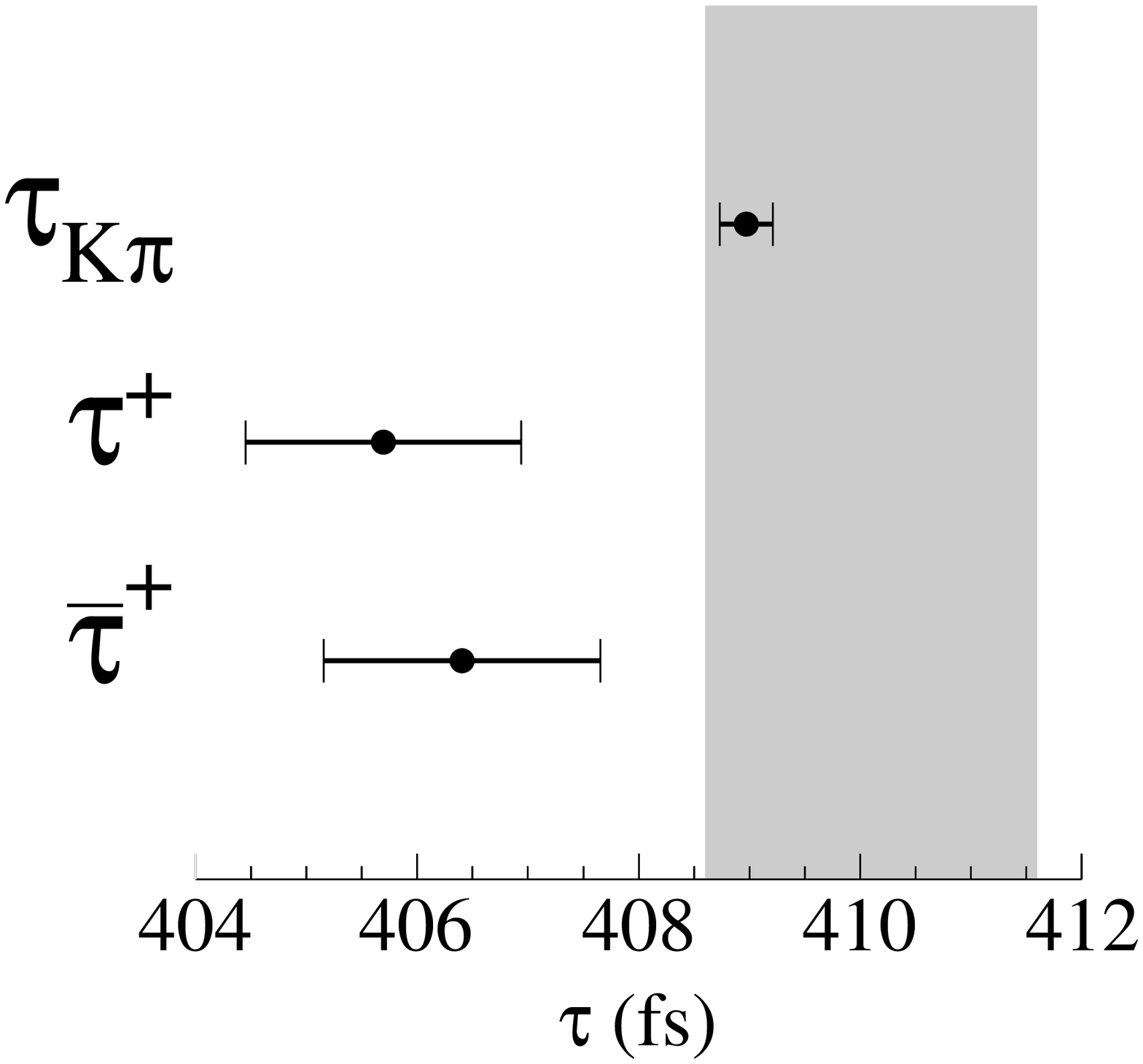}  }
\caption{Proper-time $t$ distribution for each decay mode with the fit results overlaid.
The combinatorial distribution (indicated as `Comb.' in light gray) is stacked on top of the misreconstructed-charm
distribution (indicated as `Charm' in dark gray). 
The normalized Poisson pulls for each fit are shown under each plot; ``unt'' refers to the untagged datasets. 
The bottom right plot shows the individual lifetimes (with statistical uncertainties only);
the gray band indicates the PDG \Dz lifetime $\pm1\sigma$~\cite{Nakamura:2010zzi}.}
\label{fig:DecayTime}
\end{figure}

The simultaneous fit to all events in the lifetime-fit mass region
has 20 floating parameters: the seven signal yields and three signal lifetimes;
 the yield of untagged $\KK$ combinatorial candidates; the offset \toff;
the parameters $f_{t1}$ and $f_{t2}$ characterizing the weight of each Gaussian in the signal resolution mode;
and the proper-time error scaling parameters $s_1,\ s_2,\ s_3,\ S_{\kk},\ S_{\pi\pi}$, and $S'_{\rm tag}$.
After extracting the three signal lifetimes, using their
reciprocals in the computation of $\yCP$ and $\deltaY$ as
defined in Eqs.~(\ref{eq:yCPcalc2}) and~(\ref{eq:deltaYcalc2}), respectively, we find
\begin{eqnarray*}
\yCP    &=& [0.72 \pm 0.18 \stat]\% , \\
\deltaY &=& [0.09 \pm 0.26 \stat]\% .
\end{eqnarray*}
The statistical errors are computed using the covariance matrix returned by the fit.
The lifetime-fit mass region proper-time distributions and projections of the lifetime fit
for the seven different decay modes are shown in Fig.~\ref{fig:DecayTime}.

\section{cross checks and systematics}

We have performed numerous  cross checks to search
for potential problems, in addition to quantitative studies
that yield the systematic uncertainties given in
Table~\ref{tab:SystematicVariations}, discussed below.
Initially we tested the fit model by generating large ensembles of
 datasets randomly drawn from the underlying total PDF, and observed no biases in the $\yCP$ and $\deltaY$
results obtained.
In addition, we have fit an ensemble of four simulated datasets, each equivalent in luminosity to the data,
 and found no evidence of bias in  $\yCP$ or $\deltaY$.

In fitting the data, we find that the tagged and untagged extracted lifetimes for $\KK$, and separately for $\DzRW$,
 are compatible within the statistical uncertainties.
We performed a simultaneous fit to the tagged channels, and a separate simultaneous fit to the untagged channels, and find the lifetimes
to be compatible within the statistical uncertainties.
We  repeated the fit allowing the $\KK$ and $\pipi$ final states to have separate $\tauhhp$ and $\tauhhm$
lifetimes, and  observed no statistically significant difference between the $\KK$ and $\pipi$
results. We  estimated the effects of the SVT misalignment to be negligible.

We  varied the lifetime-fit mass region width by $\pm 4$ and $\pm 2 \mevcc$.
We adopt as the systematic uncertainty half the RMS of the differences
$|\Delta[\yCP]|$ and $|\Delta[\deltaY]|$ from the nominal fit central values.
We also shifted the position of each mass region by centering each of them at
the most probable value for the signal PDF obtained in the invariant mass fits.
These systematic uncertainties are given in the first two lines of Table~\ref{tab:SystematicVariations}.

For the  untagged $\KK$ mode, the combinatorial yield is a parameter determined in the lifetime fit.
However, it is also needed to determine the signal \terr PDF. We first
use the total background yield determined from the mass fit to extract a signal \terr PDF, which is employed in an initial simultaneous 
lifetime fit. The combinatorial yield from this fit is used to construct an improved \terr signal PDF 
 and a second fit is performed (the nominal fit). We estimate the systematic error on \yCP and \deltaY associated 
with the determination of the signal \terr PDF for the untagged $\KK$ mode 
to be the difference in the values obtained from an additional iteration of the fit and the nominal fit.

We vary the nominal mistag rate of $0.2\%$ by $\pm 0.04\%$,
a $20\%$ relative variation, and find no significant change in the nominal fit values.
Instead of assuming equal fractions of $\Dz$ and $\Dzb$ in the untagged $\KK$
mode, we adopt the latest CDF result for direct $\CPV$~\cite{Aaltonen:2011se},
and find negligible change in $\yCP$ and $\deltaY$.

We rely on simulated events to determine both the PDF shapes and yields
for the misreconstructed-charm backgrounds. To account for the model
dependence, we vary the effective lifetime of these events
by $\pm 5\%$, except for the tagged $\pipi$ mode where the variation is $\pm 15\%$
due to the small number of simulated events that pass the selection criteria for this mode.
We also vary the expected misreconstructed-charm yields by $\pm 10\%$ in the
tagged channels, and $\pm 5\%$ in the untagged channels. Each variation is
simultaneously applied to all modes. These are $\gae 2 \sigma$ variations
relative to the statistical uncertainties of the simulated datasets.

We vary the yields, weighting parameters, and fitting strategy used
to obtain the  2-d lifetime PDF for combinatorial-background events in the lifetime-fit mass region
from the mass sidebands. The yields for the tagged combinatorial-background events are varied by $\sim 5\%$ in the $\pipi$ mode,
$15\%$ in tagged $\KK$, and $20\%$ in $\DzRW$.
The untagged $\DzRW$ combinatorial-background yield is varied using the value extracted from
an alternative lifetime-fit model in which the yield is allowed to vary.

The weights given to the low- and high-mass sidebands in the data in order to derive the combinatorial PDF
in the lifetime-fit mass region in data are extracted from simulated events.
They are varied by plus and minus the statistical uncertainty derived from splitting the
simulated dataset, which is equivalent to several times the nominal integrated luminosity,
into datasets that numerically match the nominal luminosity.

We also apply the variations described above for the misreconstructed-charm
background to vary the yields and shape of the PDF that describe the residual signal events in the sidebands.
This is also done for the misreconstructed-charm PDF used in the sideband fits from which the 2-d combinatorial
PDF is extracted. This yields the combinatorial PDF shape variation, which is
then used in the nominal fit, to obtain the variation reported in Table~\ref{tab:SystematicVariations}.

Finally, we vary the \terr criteria by $\pm 0.1 \ps$ from the nominal $\terr < 0.5 \ps$,
and take as the systematic uncertainty the RMS of the deviations from the nominal fit central value divided by $\sqrt{2}$.
We also consider two variations in how multiple candidates sharing one or more daughter tracks are treated.  
In the first variation, we retain all multiple candidates, if each candidate passes all the
other selection criteria. In the second variation, we reject all multiple candidates sharing one
or more daughter tracks.
We fit these datasets using the nominal fit model, and assign the largest observed
deviation from the nominal $\yCP$ and $\deltaY$ central values as the systematic uncertainty
 in Table~\ref{tab:SystematicVariations}.
The total $\yCP$ and $\deltaY$ systematic uncertainties are calculated by summing the contributions
 from all sources in quadrature, and are reported in the last row of Table~\ref{tab:SystematicVariations}.

\begin{table}[]
\begin{center}
\caption{The \yCP and \deltaY systematic uncertainties.
The total is the sum-in-quadrature of the entries in each column.}
\begin{tabular}{l|c|c}  \hline\hline
Fit Variation & $|\Delta[\yCP]|$ (\%)  & $|\Delta[\deltaY]|$ (\%)\\
\hline
mass window width         & 0.057 & 0.022    \\
mass window position      & 0.005 & 0.001    \\

untagged \kk signal \terr PDF         & 0.022 & 0.000     \\
mistag fraction             & 0.000   & 0.000      \\
untagged \kk \Dz fraction & 0.001 & 0.000     \\

charm bkgd. lifetimes     & 0.042 & 0.001    \\
charm bkgd. yields        & 0.016 & 0.000      \\

comb. yields             & 0.043 & 0.002    \\
comb. sideband weights   & 0.004 & 0.001    \\
comb. PDF shape          & 0.066 & 0.000      \\

\terr selection          & 0.052  & 0.053  \\
candidate selection      & 0.028  & 0.011  \\ 
\hline
Total                    & 0.124  & 0.058  \\
\hline\hline
\end{tabular}
\label{tab:SystematicVariations}
\end{center}
\end{table}

\section{Conclusions}

In summary, we  measured $\yCP$ and $\deltaY $ to a precision significantly better than our previous 
measurements~\cite{Aubert:2007en,Aubert:2009ai}. 
Both results are more precise than, and consistent with, the weighted average of all previous measurements~\cite{Asner:2010qj},
when the previous \babar\, results are excluded. In particular, the \yCP measurement is the most precise single measurement to date.
We obtain
\begin{eqnarray*}
\yCP     &=&  [0.72 \pm 0.18 \stat \pm 0.12 \syst]\% , \\
\deltaY  &=&  [0.09 \pm 0.26 \stat \pm 0.06 \syst]\% .
\end{eqnarray*}
We exclude the null mixing hypothesis at $3.3 \sigma$ significance,
and find no evidence for $\CPV$.
Our results are consistent with the world average value of the mixing parameter $y$
 obtained from $\Dz \to \KS h^{-}h^{+}$ (where $h=K,\pi$)~\cite{Asner:2010qj}, as expected in absence of \CPV.

The value of $\deltaY$ obtained here is consistent with our previously published result~\cite{Aubert:2007en}
when the same definition is used in both cases.
The new $\yCP$ value is consistent with our previous result~\cite{Aubert:2009ai} with a probability of 
 $\gae 2\%$, assuming that the systematics for both the old and new
measurements are fully correlated, and taking into account the fact that $\sim40\%$ of
 the events in the current sample are also present in the samples used in the previous measurements~\cite{Aubert:2007en,Aubert:2009ai}.
The results here supersede the previous \babar\, results
for these modes~\cite{Aubert:2007en,Aubert:2009ai}.

\section*{ Acknowledgments }

\input{acknowledgements.tex}

\clearpage

\onecolumngrid

\appendix

\section{Mixing Formalism and Considerations on the Role of Direct \CP Violation}
\label{app:mixing}

The time evolution of the flavor eigenstates \Dz and \Dzb is governed by the Schr\"odinger equation:

\begin{equation}
  i \frac{\partial}{\partial t}
  \left( \begin{array}{c} \Dz(t)\\
    \Dzb(t) \\
  \end{array} \right) =
  \big(\mathbf{M}-\frac{i}{2}\mathbf{\Gamma}\big)
  \left( \begin{array}{c} \Dz(t)\\
  \Dzb(t) \\
  \end{array} \right)\;. \nonumber
\end{equation}\\
The mass eigenstates $D_1$ and $D_2$ are obtained from the diagonalization of the effective Hamiltonian
$\mathcal{H}_{\rm eff} = \mathbf{M}-\frac{i}{2}\mathbf{\Gamma}$.
Under the hypothesis of \CPT conservation the two mass eigenstates can be written in terms of the flavor eigenstates as
\begin{equation}
\begin{array}{rcl}
| D_1 \rangle &=& p | \Dz \rangle + q | \Dzb \rangle \;, \\
| D_2 \rangle &=& p | \Dz \rangle - q | \Dzb \rangle \;,
\end{array}
\label{eq:qpdef}
\end{equation}
where
\begin{equation}
\left(\frac{q}{p}\right)^2 = \frac{M_{12}^*-\frac{1}{2}\Gamma_{12}^*}{M_{12}-\frac{1}{2}\Gamma_{12}} \qquad \mbox{and} \qquad  \left|p\right|^2 + \left|q\right|^2 = 1.
\label{eq:qoverpdef}
\end{equation}
We choose the positive root for $q/p$; choosing the negative one just means exchanging $D_1$ with $D_2$.
If $\CP | \Dz \rangle = +| \Dzb \rangle $, in case of no \CPV,
$D_1$ is the $\CP$-even state and $D_2$ the $\CP$-odd state.

It is traditional to quantify the size of $\DzDzb$ mixing
in terms of the parameters $x \equiv \Delta m/\Gamma$ and
$y \equiv \Delta\Gamma/2\Gamma$, where $\Delta m = m_1 - m_2$
($\Delta \Gamma = \Gamma_1 - \Gamma_2$) is the difference in
mass (width) of the states defined in Eq.~\ref{eq:qpdef}
and $\Gamma = (\Gamma_1+\Gamma_2)/2$ is the average width.
If either $x$ or $y$ is non-zero, mixing will occur.
While most Standard Model expectations for the size of both are
$\lesssim 10^{-3}$~\cite{Falk:2001hx,Nelson:1999fgSupp}, values as high as
$10^{-2}$ or even higher are predicted by certain models~\cite{Petrov:2006nc,Golowich:2007ka}.

\CP violation can manifest in $D^0$ decays in three ways:
\begin{itemize}
\item in decay, when $|A_f/\bar{A}_f| \neq 1$,
\item in mixing, when  $r_m = |q/p| \neq 1$,
\item in the interference between decays with and without mixing, when the weak phase $\phi_f$ of $\lambda_f \equiv \frac{q}{p}\frac{\bar{A}_f}{A_f}$ is different from zero,
\end{itemize}

where $A_f$ ($\bar{A}_f$) is the amplitude for $D^0$ ($\Dzb$) decaying into a final state $f$,
  $ A_f \equiv \langle f | {\mathcal H}_D | D^0 \rangle$  ($\bar{A}_f \equiv \langle f | {\mathcal H}_D | \Dzb \rangle$).

The presence of mixing alters the exponential distribution for the \Dz decay into a final state $f$. In particular we have
\begin{eqnarray}
\Gamma(\Dz(t) \to f)  & = \frac{1}{2}|A_f|^2e^{-\Gamma t}&\left[ (1 + |\lambda_f|^2)\cosh y\Gamma t + (1 - |\lambda_f|^2)\cos x\Gamma t   \right. \\ \nonumber
                      & & \left. -2\Re(\lambda_f) \sinh y\Gamma t + 2\Im(\lambda_f) \sin x \Gamma t  \right]\: ,\\
\label{fullGammaDztof}
\Gamma(\Dzb(t) \to f)  & = \frac{1}{2}|\bar{A}_f|^2e^{-\Gamma t}&\left[ (1 + |\lambda_f^{-1}|^2)\cosh y\Gamma t + (1 - |\lambda_f^{-1}|^2)\cos x\Gamma t   \right.\\ \nonumber
                      & & \left. - 2\Re(\lambda_f^{-1}) \sinh y\Gamma t + 2\Im(\lambda_f^{-1}) \sin x \Gamma t  \right].
\label{fullGammaDzbtof}
\end{eqnarray}

In this analysis we are interested in \CP-even final states ($f = h^+h^-, h=K,\pi$).
If we neglect second-order terms in $x\Gamma t$ and $ y\Gamma t$, the decay
time distributions can be treated as exponentials with effective widths~\cite{Bergmann:2000idSupp}:
\begin{eqnarray}
\Gamma(\Dz(t) \to f)  &\propto& e^{-\Gammahhp t} \quad \mbox{with} \quad \Gammahhp   = \Gamma\left[ 1 + y\ \Re(\lambda_{hh}) - x\ \Im(\lambda_{hh}) \right]\: , \label{gammahhp} \\
\Gamma(\Dzb(t) \to f) &\propto& e^{-\Gammahhm t} \quad \mbox{with} \quad \Gammahhm = \Gamma\left[ 1 + y\ \Re(\lambda_{hh}^{-1}) - x\ \Im(\lambda_{hh}^{-1})  \right].
\label{gammahhm}
\end{eqnarray}

To better understand the effects of \CP violation we introduce two more parameters, one describing \CPV in decay ($A_D^f$) and one in mixing ($A_M$):
\begin{eqnarray}
A_D^f &=& \frac{ |A_f/\bar{A}_f|^2 - |\bar{A}_{\bar{f}}/A_{\bar{f}}|^2}{|A_f/\bar{A}_f|^2 + |\bar{A}_{\bar{f}}/A_{\bar{f}}|^2}\: , \\
A_M  &=&  \frac{ r_m^2 - r_m^{-2}}{r_m^2 + r_m^{-2}}.
\end{eqnarray}
Since $f = h^+h^-$ then $f = \bar{f}$.
Noting that there is no strong phase in $\lambda_f$ since the final state is its own \CP-conjugate,
we can express $\lambda_{hh}$ in terms of $A_D^{hh}$, $A_M$ and the \CP-violating phase $\phi_{hh}$:
\begin{equation}
\lambda_{hh}  =  \left[ \frac{1-A_D^{hh}}{1+A_D^{hh}} \;  \frac{1+A_M}{1-A_M}  \right]^{1/4} e^{i\phi_{hh}}.
\end{equation}

Expanding Eqs.~\ref{gammahhp} and~\ref{gammahhm}, and retaining only terms up to first order in $A_D^{hh}$ and $A_M$, we obtain
\begin{eqnarray}
\label{gammahhpnew}
\Gammahhp &  \simeq &\Gamma\left[ 1 + (y \cos\phi_{hh} - x \sin\phi_{hh} ) +
             \frac{1}{2}(A_M-A_D^{hh})\ (y \cos\phi_{hh} - x \sin\phi_{hh} ) \right.\\ \nonumber
          &&  \left. - \frac{1}{4}A_MA_D^{hh}\ (y \cos\phi_{hh} - x \sin\phi_{hh})\right]\: ,
\end{eqnarray}

\begin{eqnarray}
\label{gammahhmnew}
\Gammahhm &  \simeq &\Gamma\left[ 1 + (y \cos\phi_{hh} + x \sin\phi_{hh} ) -
             \frac{1}{2}(A_M-A_D^{hh})\ (y \cos\phi_{hh} + x \sin\phi_{hh} ) \right.\\ \nonumber
          &&  \left. - \frac{1}{4}A_MA_D^{hh}\ (y \cos\phi_{hh} + x \sin\phi_{hh})\right].
\end{eqnarray}

Combining the widths defined above we obtain the two observables $\yCP$ and $\deltaY$, which,
in general, depend on the final state because of the \CPV parameters $A_D^{hh}$ and $\phi_{hh}$:
\begin{eqnarray}
\yCP^{hh} &=&\frac{\Gammahhp+\Gammahhm}{2\Gamma} - 1\: ,\\
\deltaY^{hh} &=& \frac{\Gammahhp-\Gammahhm}{2\Gamma} .
\label{eq:yCPcalc}
\end{eqnarray}
Other experiments characterize the \CP-violating observable as $A_{\Gamma}$,
\begin{eqnarray}
A_{\Gamma}=\frac{\Gammahhp-\Gammahhm}{\Gammahhp-\Gammahhm}.
\end{eqnarray}
The relationship between $A_{\Gamma}$, \deltaY and \yCP is
\begin{eqnarray}
\deltaY = (1+\yCP)\ A_{\Gamma}.
\end{eqnarray}

These quantities are directly related to the fundamental parameters that govern mixing and \CPV in the charm sector:
\begin{eqnarray}
\yCP^{hh}    &=& \phantom{-}y\cos\phi_{hh} - \frac{1}{2}\left[A_M+A_D^{hh}\right]\ x\sin\phi_{hh} - \frac{1}{4}A_MA_D^{hh}\ y\cos\phi_{hh}\: , \\
\deltaY^{hh} &=& - x\sin\phi_{hh} + \frac{1}{2}\left[A_M+A_D^{hh}\right]\ y\cos\phi_{hh} + \frac{1}{4}A_MA_D^{hh}\ x\sin\phi_{hh}.
\label{eq:yCPfund}
\end{eqnarray}
Both $\yCP$ and $\Delta Y$ are zero if there is no $\DzDzb$ mixing. Otherwise,
a non-zero value of \yCP implies mixing and a non-zero value of \deltaY implies \CPV.

In the charm sector, because the CKM elements involved belong to the Cabibbo submatrix,
we can assume that the weak phase $\phi_{hh}$ does not depend
on the final state: $\phi_{hh} = \phi$~\cite{Kagan:2009gb}. As stated earlier if direct \CPV has a 
 significant effect, then the values of \yCP and \deltaY depend on the final state.
In this analysis we assume that the effect of direct \CPV is negligible in the
decays to \CP eigenstates; {\it i.e.}, we assume $\Gamma_{KK}^+ = \Gamma^+_{\pi\pi}$
(and $\overline{\Gamma}^+_{KK} = \overline{\Gamma}^+_{\pi\pi}$).
In Eqs.~\ref{gammahhpnew} and~\ref{gammahhmnew} this means neglecting
the linear terms in $A_D^{hh}$.
Assuming that $A_D^{hh}$ and $y$ are both $\mathcal{O}(1\%)$ and $\phi_{hh} = 0$,
the neglected term is  $\mathcal{O}(10^{-4})$, beyond any current experimental sensitivity.

Under the above assumptions, Eqs.~\ref{gammahhpnew} and~\ref{gammahhmnew} become
\begin{eqnarray}
\label{gammahhpnew2}
\Gamma^+ &  \simeq &\Gamma\left[ 1 + (y \cos\phi - x \sin\phi ) +
             \frac{A_M}{2}\ (y \cos\phi - x \sin\phi )\right] \: ,\\
\label{gammahhmnew2}
\overline{\Gamma}^+ &  \simeq &\Gamma\left[ 1 + (y \cos\phi + x \sin\phi ) -
             \frac{A_M}{2}\ (y \cos\phi + x \sin\phi)\right].
\end{eqnarray}

Inserting Eqs.~\ref{gammahhpnew2} and~\ref{gammahhmnew2} into Eqs.~\ref{eq:yCPcalc2} and~\ref{eq:deltaYcalc2} yields
\begin{eqnarray}
\yCP &=&  \phantom{-}y\cos\phi - \frac{A_M}{2} x\sin\phi \: ,\\
\deltaY &=& -x\sin\phi + \frac{A_M}{2} y\cos\phi.
\label{eq:yCPfund2}
\end{eqnarray}
From the experimental point of view, we measure three lifetimes instead of the partial widths: 
\begin{itemize}
\item \tauhhp for the $\Dz \to  \KmKp,\ \pimpip$ decays,
\item  \tauhhm for the $\Dzb \to  \KmKp,\ \pimpip$ decays,
\item \tauKpi for the $\Dz$ (and \Dzb) $ \to  \DzRW$ decays (the Cabibbo favored $\Km\pip$ and the doubly Cabibbo suppressed $\Kp\pim$ decays are collected in the same sample),
\end{itemize}
and use their inverse to compute \yCP and \deltaY.

The measured observables constrain the parameters that govern mixing and indirect \CPV\ in the charm sector.
\section{Signal Lifetime PDFs}
\label{app:pdf}

The explicit form of the signal lifetime PDFs based on the prototype PDFs presented in the main text are given below:

\begin{eqnarray*}
{\cal P}_{\pi\pi}^{\Dstp}(t,\terr) & = & (1-f^+_{\rm tag}){\cal R}_{\pi\pi}^{\rm tag}(t,\terr;\Spipi \Stag s_i,\toff,\tauhhp) + f^+_{\rm tag}{\cal R}_{\pi\pi}^{\rm tag}(t,\terr;\Spipi \Stag s_i,\toff,\tauhhm)\, , \\ \\
{\cal P}_{\pi\pi}^{\Dstm}(t,\terr) & = & (1-f^-_{\rm tag}){\cal R}_{\pi\pi}^{\rm tag}(t,\terr;\Spipi \Stag s_i,\toff,\tauhhm) + f^-_{\rm tag}{\cal R}_{\pi\pi}^{\rm tag}(t,\terr;\Spipi \Stag s_i,\toff, \tauhhp)\, ,  \\\\
{\cal P}_{KK}^{\Dstp}(t,\terr) & = & (1-f^+_{\rm tag}){\cal R}_{KK}^{\rm tag}(t,\terr;\SKK \Stag s_i,\toff,\tauhhp) + f^+_{\rm tag}{\cal R}_{KK}^{\rm tag}(t,\terr;\SKK \Stag s_i,\toff,\tauhhm)\, , \nonumber \\  \\
{\cal P}_{KK}^{\Dstm}(t,\terr) & = &(1-f^-_{\rm tag}){\cal R}_{KK}^{\rm tag}(t,\terr;\SKK \Stag s_i,\toff,\tauhhm) + f^-_{\rm tag}{\cal R}_{KK}^{\rm tag}(t,\terr;\SKK \Stag s_i,\toff,\tauhhp)\, , \nonumber \\  \\
{\cal P}_{K\pi}^{\Dst\pm}(t,\terr) & = &{\cal R}_{K\pi}^{\rm tag}(t,\terr;\SKpi \Stag s_i,\toff,\tauKpi)\, , \\\\
{\cal P}_{KK}^{\rm unt}(t,\terr) & = &(1-f_{\Dz}){\cal R}_{KK}^{\rm unt}(t,\terr;\SKK \Sunt s_i,\toff,\tauhhm) + f_{\Dz}{\cal R}_{KK}^{\rm unt}(t,\terr;\SKK \Sunt s_i,\toff,\tauhhp)\, ,  \nonumber \\\\
{\cal P}_{K\pi}^{\rm unt}(t,\terr) & = &{\cal R}_{K\pi}^{\rm unt}(t,\terr;\SKpi \Sunt s_i,\toff,\tauKpi)\, , 
\end{eqnarray*}

where $f^{\pm}_{\rm tag} = 0.2\%$, $f_{\Dz} = 0.5$ and $\SKpi = \Sunt =1$ are fixed in the nominal fit.

\end{document}

%% file: authors_apr2012.tex
%
\author{J.~P.~Lees}
\author{V.~Poireau}
\author{V.~Tisserand}
\affiliation{Laboratoire d'Annecy-le-Vieux de Physique des Particules (LAPP), Universit\'e de Savoie, CNRS/IN2P3,  F-74941 Annecy-Le-Vieux, France}
\author{J.~Garra~Tico}
\author{E.~Grauges}
\affiliation{Universitat de Barcelona, Facultat de Fisica, Departament ECM, E-08028 Barcelona, Spain }
\author{A.~Palano$^{ab}$ }
\affiliation{INFN Sezione di Bari$^{a}$; Dipartimento di Fisica, Universit\`a di Bari$^{b}$, I-70126 Bari, Italy }
\author{G.~Eigen}
\author{B.~Stugu}
\affiliation{University of Bergen, Institute of Physics, N-5007 Bergen, Norway }
\author{D.~N.~Brown}
\author{L.~T.~Kerth}
\author{Yu.~G.~Kolomensky}
\author{G.~Lynch}
\affiliation{Lawrence Berkeley National Laboratory and University of California, Berkeley, California 94720, USA }
% ...
\collaboration{The \babar\ Collaboration}
\noaffiliation

%% file: acknowledgements.tex
We are grateful for the 
extraordinary contributions of our \pep2\ colleagues in
achieving the excellent luminosity and machine conditions
that have made this work possible.
The success of this project also relies critically on the 
expertise and dedication of the computing organizations that 
support \babar.
The collaborating institutions wish to thank 
SLAC for its support and the kind hospitality extended to them. 
This work is supported by the
US Department of Energy
and National Science Foundation, the
Natural Sciences and Engineering Research Council (Canada),
the Commissariat \`a l'Energie Atomique and
Institut National de Physique Nucl\'eaire et de Physique des Particules
(France), the
Bundesministerium f\"ur Bildung und Forschung and
Deutsche Forschungsgemeinschaft
(Germany), the
Istituto Nazionale di Fisica Nucleare (Italy),
the Foundation for Fundamental Research on Matter (The Netherlands),
the Research Council of Norway, the
Ministry of Education and Science of the Russian Federation, 
Ministerio de Ciencia e Innovaci\'on (Spain), and the
Science and Technology Facilities Council (United Kingdom).
Individuals have received support from 
the Marie-Curie IEF program (European Union) and the A. P. Sloan Foundation (USA).